\documentclass[12pt, titlepage]{article}

\usepackage{amssymb}
\usepackage{amsmath}
\usepackage{graphicx}
\usepackage[english, polish]{babel}
\begin{document}

\title{The FEM approach to the 3D electrodiffusion on 'meshes' optimized with the Metropolis algorithm}

\author{I. D. Kosi\'nska\\Wroclaw University of Technology\\Institute of Biomedical Engineering and Instrumentation\\
\selectlanguage{polish}
 Wybrze\.ze Wyspia\'nskiego 27, 50-370 Wroc"law, Poland } 


\date{}
\maketitle

\selectlanguage{english}

\begin{abstract}

The presented article contains a 3D mesh generation routine optimized with the Metropolis algorithm. The procedure enables to produce meshes of a prescribed volume $V_0$ of elements. The finite volume meshes are used with the Finite Element approach. The FEM analysis enables to deal with a set of coupled nonlinear differential equations that describes the electrodiffusional problem. Mesh quality and accuracy of FEM solutions are also examined. High quality of FEM type space--dependent approximation and correctness of discrete approximation in time are ensured by finding solutions to the 3D Laplace problem and to the 3D diffusion equation, respectively. Their comparison with analytical solutions confirms accuracy of obtained approximations.

\end{abstract}





\section{Introduction}

One from the most important physical processes is electrodiffusion. It describes both diffusional motion of mass and charge flow due to applied electric field. The electric potential distribution is govern by the Poisson equation and total transport of particles is given in terms of the continuity equation \cite{kubo}. The significance of this equation is broadly described in existing physical, chemical and biological literature \cite{kampen} and lots of scientific articles, particularly, those which concern properties of nano and micro transport.\\
Mathematically, equations of electrodiffusion constitute a set of coupled nonlinear equations where the Laplace operator \cite{feynman, evans, courant} appears together with the first order partial time derivative. The Laplace operator is the basic operator met in many physical situations. Thus the first step to deal with the electrodiffusional problem is to approximate the solution of the Laplace equation with help of the Finite Elements Method. Practically, it means that an appropriate mesh should be designed for a prescribed 3D domain. The mesh must fit well to the physical conditions like e. g. symmetry of the problem. Therefore, different mesh shapes could be desired (spherical, cylindrical, conical or cubic) up to problem. After having accurate basic spatial solutions on appropriate meshes, the problem should be extended to the time--dependent case of the diffusion equation by finding discrete approximation in time. It could be done by means of truncated Taylor series or other single--step procedures like the Crank--Nicolson scheme \cite{zienkiewicz, crank} or the Gurtin's approach to finite element approximation in terms of variational principle \cite{gurtin}. From now, further extension of above-presented computations involving non-linear terms could be easily implemented and numerically solved using the Newton's method \cite{kelly}.             

\section{Equation of electrodiffusion}
The equation of electrodiffusion \cite{kubo} has the form
\begin{eqnarray}
\nabla\cdot J_i + \frac{\partial n_i}{\partial t} = 0
\label{electrodiffusion}
\\
J_i = -\tilde{D}_i\left(\nabla n_i + \frac{z_i e n_i}{kT}\nabla \phi\right) \nonumber
\end{eqnarray}
where $n_i$ is a number of $i$-th ions, $\phi$ - electric potential,
$\tilde{D}_i$ - the diffusion coefficient of $i$-th particles, $k_B$ - Boltzmann constant, $T$ - temperature, $z_i$ - valence of the $i$-th kind of ions, $e$ - electric charge. To find the electric potential $\phi$ the Poisson equation must be solved
\begin{equation}
\nabla^T(\epsilon_0\epsilon\nabla \phi) + \rho = 0
\label{Poisson}
\end{equation}
where $\rho = \sum_i \rho_i$, $\rho_i = z_i e n_i$. Thus equations (\ref{electrodiffusion}) and (\ref{Poisson}) both constitute the system of coupled equations.
\subsection{FEM approach}
Next, they can be solved numerically using the Finite Element Method \cite{zienkiewicz}
where the problem is represented as
\begin{equation}
\begin{array}{c}
\vspace{10pt}
\displaystyle \int_\Omega \mathbf{v}^T \mathbf{A(u)} d\Omega \equiv \int_\Omega [v_1A_1(\mathbf{u}) + v_2A_2(\mathbf{u}) + v_3A_3(\mathbf{u})]d\Omega = 0 \\
\displaystyle \int_\Gamma \mathbf{\tilde{v}}^T \mathbf{B(u)} d\Gamma \equiv \int_\Gamma [\tilde{v}_1 B_1(\mathbf{u}) + \tilde{v}_2 B_2(\mathbf{u}) + \tilde{v}_3 B_3(\mathbf{u}) ] d\Gamma = 0
\end{array}
\end{equation}
where
\begin{equation}
\mathbf{u} = \left[ 
\begin{array}{c}
n_+\\
n_-\\
\phi
\end{array}
\right]
\end{equation}
and $\mathbf{v}$ and $\mathbf{\tilde{v}}$ are sets of arbitrary functions equal in number to the number of equations (or components of $\mathbf{u}$) involved. $A_1(\mathbf{u}), A_2(\mathbf{u})$ and $A_3(\mathbf{u})$ are given by the following formulas, respectively
\begin{equation}
\begin{array}{c}
\vspace{10pt}
A_1(\mathbf{u})  = \nabla^T \left(-k_+\mathbf{u}^T 
\left[
\begin{array}{c}
1   \\
0 \\
0 
\end{array}
\right]
\nabla \mathbf{u}^T\left[
\begin{array}{c}
0\\
0\\
1
\end{array}
\right] - \tilde{D_+}
  \nabla\mathbf{u}^T 
\left[
\begin{array}{c}
1\\
0\\
0
\end{array}
\right]
\right) + [1, ~0, ~0] \displaystyle\frac{\partial}{\partial t}\mathbf{u}\\
\vspace{10pt}
A_2(\mathbf{u})  = \nabla^T \left(-k_-\mathbf{u}^T 
\left[
\begin{array}{c}
0 \\
1 \\
0 
\end{array}
\right]
\nabla \mathbf{u}^T\left[
\begin{array}{c}
0\\
0\\
1
\end{array}
\right] - \tilde{D}_-
  \nabla\mathbf{u}^T 
\left[
\begin{array}{c}
0\\
1\\
0
\end{array}
\right]
\right) + [0,~1,~0] \displaystyle\frac{\partial}{\partial t}\mathbf{u}\\
A_3(\mathbf{u}) = \epsilon_0\epsilon\nabla^T\nabla\mathbf{u}^T 
\left[
\begin{array}{c}
0\\
0\\
1
\end{array}
\right] + [1, ~-1, ~0]ze\mathbf{u}.
\end{array}
\end{equation}
where $\displaystyle k_i = \frac{D_iz_ie}{k_BT}$, $i=\left\{ +, -\right\}$.
An expression $\mathbf{B(u)}$ gives the boundary conditions on $\Gamma$, however, we choose a \emph{forced} type of boundary conditions on $\Gamma$ i. e.
\begin{equation}
\begin{array}{c}
\phi - \phi_{boun} = 0 \\
n_i - n_{i, ~boun} = 0. 
\end{array}
\label{boun}
\end{equation}
Let us substitute $\mathbf{u} \approx  
\sum_a N_a \mathbf{I} \left[
\begin{array}{c}
\tilde{n^a}_+\\
\tilde{n^a}_-\\
\tilde{\phi^a}
\end{array}
\right]
= \mathbf{N}\mathbf{\tilde{u}} $ and
put $\mathbf{v} = \sum_b w_b \mathbf{I} \delta\mathbf{\tilde{u}_b}$ where $w_b = N_b$. In that way, we end up with the Galerkin formulation of the problem. 

\subsubsection{Discrete approximation in time}
In turn, we can approximate the nodal electric potential 
\begin{equation}
\phi(x,y,z,t) = \sum_a N_a(x,y,z)\tilde{\phi^a}(t)
\end{equation} 
and number of particles 
\begin{equation}
n_i(x,y,z,t)= \sum_a N_a(x,y,z)\tilde{n_i^a}(t)
\end{equation}
at a time $t_n$ by
\begin{eqnarray}
\tilde{\phi}^a(t_n) \approx \tilde{\phi}^a_n\\
\tilde{n}_i^a(t_n) \approx \tilde{n}^a_{i,n}
\end{eqnarray}
and taking advantage from an evaluation of $\mathbf{\tilde{u}}_n = \left[
\begin{array}{c}
\tilde{n}_{+,n}\\
\tilde{n}_{-,n}\\
\mathbf{\tilde{\phi}}_n
\end{array}
\right]$ in the Taylor series we obtain
\begin{equation}
\mathbf{\tilde{u}}_{n} \approx \mathbf{\tilde{u}}_{n-1} + \Delta t \mathbf{\dot{\tilde{u}}}_{n-1} + \beta(\Delta t)^2 \mathbf{\ddot{\tilde{u}}}_{n-1} + O(\Delta t^3)
\label{discrete_in_time}
\end{equation}
where $\beta$ takes values from $[0, 1]$ and $\Delta t$ denotes time step. After incorporating it into a general form of time--dependent equations $A_{\{1,2\}}(\mathbf{u})$
\begin{equation}
K(\mathbf{u}) + \mathbf{C}\dot{\mathbf{u}} = 0
\end{equation}
where $K(\mathbf{u})$ represents these parts of $A_{\{1,2\}}(\mathbf{u})$ with a space--dependent operator we get time approximation for a given node
\begin{equation}
\displaystyle K(N_a\mathbf{I}\mathbf{u}^a_{n}) + \mathbf{C} N_a\mathbf{I}\left\{\frac{1}{\beta\Delta t}\left(\mathbf{u}^a_{n} - \mathbf{u}^a_{n-1}\right) - \frac{1 - \beta}{\beta} \mathbf{\dot{u}}^a_{n-1}\right\} = 0. 
\end{equation}
When $\beta = 1$ an approximate solution to the semi-discrete equations at each time $t_n$ is given by the Euler ,,backward'' scheme
\begin{equation}
K(\mathbf{N\tilde{u}}) + \mathbf{C}  \mathbf{N}\frac{1}{\Delta t}\mathbf{\tilde{u}_n} =  \mathbf{C} \mathbf{N}\frac{1}{\Delta t}\mathbf{\tilde{u}_{n-1}} 
\end{equation}
otherwise, the expression for $u^a_{n}$ is as follows
\begin{equation}
\displaystyle\mathbf{C}  N_a\mathbf{I}\frac{1}{\beta\Delta t} \mathbf{u}^a_{n}  + K(N_a\mathbf{I}\mathbf{u}^a_{n}) -\mathbf{C}  N_a \mathbf{I}\frac{1}{\beta\Delta t} \mathbf{u}^a_{n-1} + \frac{1 - \beta}{\beta} K(N_a \mathbf{I}\mathbf{u}^a_{n-1}) = 0.  
\end{equation}

\subsubsection{Space--dependent term}
After integration $NK(\mathbf{N\tilde{u}})$ by parts, one obtains a mixed set of linear and nonlinear equations
\begin{equation}
\begin{array}{c}
\vspace{10pt}
\displaystyle \int_\Omega(\nabla N_b)^Tk_+\left(\sum_a N_a\tilde{n}^{+,a}\right)
\nabla \left(\sum_{c} N_{c}\tilde{\phi}^c\right) d\Omega + \int_\Omega\left( \nabla N_b \right)^T\tilde{D}_+\nabla\left( \sum_a N_a \tilde{n}^{+, a}\right)d\Omega \\
\vspace{10pt}
\displaystyle + \frac{1}{\Delta t}
\int_\Omega N^b \sum_a N_a \left( \tilde{n}^{+, a}_n - \tilde{n}^{+, a}_{n-1} \right) d\Omega 
= 0 \\
\vspace{10pt}
\displaystyle \int_\Omega(\nabla N_b)^Tk_- \left(\sum_a N_a\tilde{n}^{-, a}\right)
\nabla \left(\sum_{c} N_{c}\tilde{\phi}^c\right) d\Omega + \int_\Omega\left( \nabla N_b \right)^T\tilde{D}_-\nabla\left( \sum_a N_a \tilde{n}^{-, a}\right)d\Omega \\
\vspace{10pt}
\displaystyle + \frac{1}{\Delta t}
 \int_\Omega N^b \sum_a N_a \left( \tilde{n}^{-,a}_n - \tilde{n}^{-,a}_{n-1} \right) d\Omega 
= 0 \\
\vspace{10pt}
\displaystyle - \int_\Omega\left(
\nabla N_b
\right)^T\epsilon_0\epsilon\nabla\left(\sum_a N_a \tilde{\phi}^a \right) d\Omega
\displaystyle + z_+e\int_\Omega N^b \left(\sum_a N_a \tilde{n}^{+,a} \right)d\Omega \\
\vspace{10pt}
\displaystyle + z_-e\int_\Omega N^b \left(\sum_a N_a \tilde{n}^{-,a} \right)d\Omega = 0\\
\quad b = 1,\dots,M
\label{omega}
\end{array}
\end{equation}
and a corresponding set of boundary terms for $\phi$ and $n^{i}$ where $i=\left\{ +, -\right\}$
\begin{equation}
\begin{array}{c}
\vspace{10pt}
- k_{i}\displaystyle\oint_\Gamma N^b \frac{\partial}{\partial n}\left\{\left(\sum_a N_a\tilde{n}^{i,a}\right)
\nabla \left(\sum_{c} N_{c}\tilde{\phi}^c\right)\right\}  d\Gamma
- \tilde{D}_{i}\displaystyle\oint_\Gamma N^b \frac{\partial}{\partial n}\left\{\nabla\left(\sum_a N_a\tilde{n}^{i,a}\right)\right\} d\Gamma
\\
+ \epsilon\epsilon_0\displaystyle\oint_\Gamma N^b\frac{\partial}{\partial n}\left\{\nabla\left(\sum_a N_a\tilde{\phi}^a\right)\right\} d\Gamma = 0  \quad b = 1,\dots,M
\label{gamma}
\end{array} 
\end{equation}
where $\displaystyle\frac{\partial}{\partial n}$ denotes derivative normal to $\Gamma$. Presented--above spatially temporal discretization has been done for the case with $\beta =1$ at each node. If the \emph{forced} boundary conditions (see Eq.~(\ref{boun})) are imposed on $\Gamma_\phi$ and $\Gamma_n$, respectively, then all terms in Eq.~(\ref{gamma}) can be neglected by restricting the choice of $N_b$ functions to those which equal 0 on $\Gamma$. Let's denote integrals from Eq.~(\ref{omega}) as
\begin{equation}
\begin{array}{c}
\vspace{10pt}
\displaystyle K_{ac}^b = \int_\Omega(\nabla N_b)^T N_a
\nabla N_{c} d\Omega = \sum_e \int_{\Omega^e}(\nabla N_b)^T N_a
\nabla N_{c} d\Omega^e = \sum_e K_{ac}^{b, e} \\
\vspace{10pt}
\displaystyle \tilde{K}_a^b = \int_\Omega\left( \nabla N_b \right)^T\nabla N_a d\Omega =  \sum_e \int_{\Omega^e}\left( \nabla N_b \right)^T\nabla N_a d\Omega^e = \sum_e \tilde{K}_a^{b, e}\\
\displaystyle K_a^b = \int_\Omega N_a N^b d\Omega =  \sum_e \int_{\Omega^e} N_a N^b d\Omega^e= \sum_e K_{a}^{b,e}.
\label{element_int}
\end{array}
\end{equation}
where $\sum_e$ with $e = 1,\dots, E$ denotes sum over elements. In 3D space tetrahedral elements seem to be a natural choice of finite volume elements. Then indices $a,b,c$ take four values each (an element has four nodes) from the set of $1,\dots,M$ values. 
\subsection{Tetrahedral elements}
For tetrahedral linear elements shape functions $N_{i}$ can be assumed as equal area coordinates $L_i$ given by the formula
\begin{equation}
\displaystyle L_{i} = \frac{a_i + b_i x + c_i y + d_i z}{6V^e}, \quad i=1,2,3,4 
\end{equation}
where $V^e$ is a volume of tetrahedron. The following integration formula can be useful
\begin{equation}
\displaystyle \int_{\Omega^e} L^\alpha_1L^\beta_2L^\mu_3L^\nu_4 dxdydz = \frac{\alpha!\beta!\mu!\nu!}{(\alpha+\beta+\mu+\nu+3)!}6V^e
\end{equation}
in order to calculate integrals $K^{b,e}_{ac}$, $\tilde{K}^{b,e}_{a}$ and $K^{b,e}_{a}$. Shape functions for linear elements are $N_a = L_a$ for $a = 1,2,3,4$. This gives  
\begin{equation}
\begin{array}{c}
\vspace{10pt}
\displaystyle K^{b,e}_{a,c} = \int_{\Omega^e}(\nabla L^b)^T L_a\nabla L_{c} d\Omega^e 
= \frac{1}{4!6V^e}[b^b, c^b, d^b] [b_c, c_c, d_c]^T =\\
\vspace{10pt}
\displaystyle \frac{1}{4!6V^e}
\left( 
b^bb_c + c^bc_c + d^bd_c
\right)\\
\vspace{10pt}
\displaystyle \tilde{K}^{b,e}_{a} = \int_{\Omega^e}\left( \nabla L^b \right)^T\nabla L_a d\Omega^e = 
\frac{1}{36V^e}
\left( 
b^bb_a + c^bc_a + d^bd_a
\right)\\
\vspace{10pt}
\displaystyle K^{b,e}_a = \int_{\Omega^e} L_a L^b d\Omega^e = \frac{6V^e}{5!} 
~~\rm{when}~~ a\ne b \\
\displaystyle K^{a,e}_a = \int_{\Omega^e} L_a L^a d\Omega^e = \frac{12V^e}{5!}.
\label{integrals}
\end{array}
\end{equation} 
Finally, we have
\begin{equation}
\begin{array}{c}
\vspace{10pt}
\displaystyle  \bigg\{ k_+\left(\sum_e K^{b,e}_{ac}\right) \tilde{\phi}^c + \tilde{D}_+ \sum_e \tilde{K}^{b,e}_{a} + \frac{1}{\Delta t}\sum_e K^{b,e}_{a} \bigg\}\tilde{n}^{+,a}  = f^{+,b},\\
\vspace{10pt}
\displaystyle \bigg\{ k_-\left(\sum_e K^{b,e}_{ac}\right)\tilde{\phi}^c  + \tilde{D}_- \sum_e \tilde{K}^{b,e}_{a} + \frac{1}{\Delta t}\sum_e K^{b,e}_{a} \bigg\} \tilde{n}^{-,a}  = f^{-,b},\\
\vspace{10pt}
\displaystyle \epsilon_0\epsilon\tilde{\phi}^a \sum_e\tilde{K}^{b,e}_{a}
= |z_i|e \sum_e K^{b,e}_{a} \left( \tilde{n}^{+,a} - \tilde{n}^{-,a} \right) \\ {\rm where} \quad a,b,c = 1,\dots, M
\label{pnp_final}
\end{array}
\end{equation}
where $\displaystyle f^{i,b} = \frac{1}{\Delta t}\left(\sum_e K^{b,e}_{a}\right)\tilde{n}^{i, a}_{n-1} $ with $i={+,-}$ and only these nodes $a, b$ and $c$ that participate in the particular element $e$ can give a non--zero contribution to the sums of the general type $\sum_e K^{e}$.\\
Let's assume that all values of $\tilde{n}^a$ are known at a time $t_n$. Then we can solve the third equation in (\ref{pnp_final}) obtaining the result
\begin{equation}
\displaystyle \tilde{\phi}^a = \frac{|z_i|e}{\epsilon_0\epsilon}\left\{\sum_e\tilde{K}^{b,e}_{a}\right\}^{-1} \left(\sum_e K^{b,e}_{a}\right) \left(\tilde{n}^{+,a} - \tilde{n}^{-,a} \right) 
\end{equation}
where $\displaystyle \left\{\sum_e\tilde{K}^{b,e}_{a}\right\}^{-1}$ denotes elements of the inverse matrix to $\tilde{K}$. After substitution the solution for $\tilde{\phi}$ to Eq.~(\ref{pnp_final}) we get
\begin{equation}
\begin{array}{c}
\vspace{10pt}
\displaystyle  \tilde{n}^{i,a} \Bigg\{\frac{k_i ze}{\epsilon_0\epsilon}\left\{\sum_e\tilde{K}^{b,e}_{c}\right\}^{-1} \left(\sum_e K^{b,e}_{c}\right) \left( \tilde{n}^{+,c} 
- \tilde{n}^{-,c} \right)  \left(\sum_e K^{b,e}_{ac}\right) \\
\displaystyle
 + \tilde{D}_i \sum_e \tilde{K}^{b,e}_{a} + \frac{1}{\Delta t}\sum_e K^{b,e}_{a} \Bigg\} - f^{i,b} = 0.
\end{array}
\end{equation}
where $i={+,-}$. 
\subsubsection{Newton's method}
The above written set of equations is of a nonlinear type. Let's denote all of them as
\begin{equation}
F(\tilde{n}^1, \dots, \tilde{n}^M) = 0
\end{equation}
Thus to solve it we have to employ the iterative Newton's method \cite{kelly}. It means that we have to start from an initial guess of $\{ \tilde{n}^i_0 \}^M_{i=1}$ values.
And during next iterations for $k=0, 1, \dots$
\begin{equation}
\tilde{n}_{k+1} = \tilde{n}_k - \left\{ F^{'}(\tilde{n}^1_k, \dots, \tilde{n}^M_k) \right\}^{-1} F(\tilde{n}^1_k, \dots, \tilde{n}^M_k)
\end{equation}
the solution should be achieved. $F^{'}(\tilde{n}^1_k, \dots, \tilde{n}^M_k)$ denotes the following matrix of partial derivatives:
\begin{equation}
F^{'}(\tilde{n}^1_k, \dots, \tilde{n}^M_k) = \left[
\begin{array}{ccc}
\vspace{10pt}
\displaystyle\frac{\partial F^1}{\partial n^1_k} & \displaystyle\dots & \displaystyle\frac{\partial F^1}{\partial n^M_k}\\
\vspace{10pt}
\vdots & \dots & \vdots\\
\displaystyle\frac{\partial F^M}{\partial n^1_k} & \dots & \displaystyle\frac{\partial F^M}{\partial n^M_k}
\end{array}
\right]
\end{equation}
where
\begin{eqnarray}
\vspace{10pt}
\displaystyle\frac{\partial F^b}{\partial n^a_k} &=& \displaystyle\Bigg\{\frac{kze}{\epsilon_0\epsilon}\left\{\sum_e\tilde{K}^{b,e}_{c}\right\}^{-1} \sum_e K^{b,e}_{c} \tilde{n}^c_k \sum_e K^{b,e}_{ac} 
 + \tilde{D} \sum_e \tilde{K}^{b,e}_{a} \\
&+& \frac{1}{\Delta t}\sum_e K^{b,e}_{a} \bigg\}
+ \displaystyle 
\frac{2kze}{\epsilon_0\epsilon}
\left\{\sum_e\tilde{K}^{b,e}_{c}\right\}^{-1}
\sum_e K^{b,e}_{c} \tilde{n}^a_k \sum_e K^{b,e}_{ac} \nonumber 
\end{eqnarray}
where $c \ne a$.

\section{Diffusion equation}
Putting $k=0$ in the equation of electrodiffusion we neglect the electrostatic term. It leads to the following equation describing diffusion in $\Re^n$ \cite{evans, courant}
\begin{equation}
\left\{
\begin{array}{ll}
u_t - D\Delta u = 0 & {\rm in} ~~\Re^n \times (0, \infty),\\
u = g & {\rm on} ~~\Re^n \times \{t=0\},
\end{array}
\right.
\label{diffusion}
\end{equation}
where $D$ denotes a diffusion coefficient. This kind of equation represents an initial value problem. Assuming that the considered domain $\Omega$ in $\Re^3$ is of a cubic type $[x_{min}, x_{max}]\times[y_{min}, y_{max}]\times[z_{min}, z_{max}]$ let us take $u = 0$ as a boundary condition. Now we seek a solution of the equation (\ref{diffusion}) which satisfies this boundary condition and prescribed initial condition at the time $t = 0$. The solution of the equation is approximated by the triple sum
\begin{equation}
u(x,y,z,t) = \displaystyle\sum_{k_x = 1}^\infty\sum_{k_y = 1}^\infty\sum_{k_z = 1}^\infty
v_{0,k_x, k_y, k_z}e^{-(k_x^2 + k_y^2 + k_z^2)D t}\sin(k_xx)\sin(k_yy)\sin(k_zz),
\end{equation}
where $v_{0,k_x, k_y, k_z}$ are unknown coefficients that must be determinated from the initial condition:
\begin{equation}
g = u(\cdot, 0) = \displaystyle\sum_{k_x = 1}^\infty\sum_{k_y = 1}^\infty\sum_{k_z = 1}^\infty
 v_{0,k_x, k_y, k_z}\sin(k_xx)\sin(k_yy)\sin(k_zz).
\end{equation}
In the case of the domain being $[0~ \pi]\times[0~ \pi]\times[0~ \pi]$ and $g=const$ the solution has the form
\begin{equation}
u(x,y,z,t) = \displaystyle \frac{64g}{\pi^3}\sum_{k_x, k_y, k_z = 1, 3, \dots}^\infty
\frac{1}{k_xk_yk_z}e^{-(k_x^2 + k_y^2 + k_z^2)D t}\sin(k_xx)\sin(k_yy)\sin(k_zz).
\end{equation}
In the case of cylindrical domain defined by $r \in [0, r_0]$, $\theta\in[0,2\pi)$ and $z\in [0, \pi]$, and with the boundary condition of the form $u = 0$ the solution of Eq.~(\ref{diffusion}) can be expanded in an absolutely and uniformly convergent series of the form
\begin{equation}
\displaystyle u(r, \theta, z, t) = \sum_{n=0}^\infty \sum_{m, k_z = 1}^\infty J_n(k_{n,m}r/r_0)\left(a_{n,m}\cos (n \theta) + b_{n,m}\sin(n\theta) \right)\sin(k_z z) e^{-((k_{n,m}/r_0)^2 + k_z^2)Dt}
\end{equation}
where $k_{n,m}$ are the zeros of the Bessel functions and $a_{n,m}, b_{n,m}$ are constants that must be found from the initial condition by making use of the \emph{orthogonality relation} for the trigonometric and Bessel functions.

\section{Laplace equation}

On the other hand, assuming that the time derivative in the diffusion equation (\ref{diffusion}) equals 0 and putting $D=1$ we end up with the boundary value problem of the Laplace type \cite{evans,courant}
\begin{equation}
\left\{
\begin{array}{ll}
\Delta u = 0 & {\rm in} ~~\Omega,\\
u = g & {\rm on} ~~\partial \Omega.
\end{array}
\right.
\label{laplace}
\end{equation}
Let's consider $\Omega$ being a cubic domain i. e. $[0~ \pi]\times[0~ \pi]\times[0~ \pi]$. And for the $g$ function equals 0 everything on $\partial\Omega$ apart from $g(x = \pi, y,z) = \phi_0$ one can approximate the exact solution by
\begin{equation}
\phi(x,y,z) = \frac{16\phi_0}{\pi^2}\sum_{n,m=1, 3, \dots}^\infty \frac{\sinh{(\sqrt{n^2+m^2}x)}\sin{(ny)}\sin{(mz)}
}{nm\sinh{(\sqrt{n^2 + m^2}\pi)}}
\end{equation}
For $\phi(x,y,z) = v(r)$ where $r = (x^2 + y^2 + z^2)^{1/2}$ the Laplace equation has the solution defined in $\Re^3$ for $r \ne 0$
\begin{equation}
\phi(x,y,z) = \frac{1}{3\alpha(3)}\frac{1}{(x^2 + y^2 + z^2)^{1/2}}
\end{equation} 
where $\alpha(3)$ denotes volume of B(0,1) in $\Re^3$ and equals $\displaystyle\frac{\pi^{3/2}}{\Gamma(3/2 + 1)}$.

\section{Three-dimensional mesh generation}

Below are listed a few technical remarks referring to the mesh generation routine applied to obtain a designed 3D mesh.

\subsection{Initial mesh}

An initial mesh is built on the basis of main surface nodes (\textit{outer} nodes) which define a figure's shape. The whole figure is considered as divided into perpendicular to $z$-axis layers. Thus the \textit{outer} nodes are distributed on the edges of layers. In the center of each layer and also in the middle between two layers are located inner nodes. They are connected with \textit{outer} nodes creating in this way the main figure's construction. Initial mesh elements obtained in such a manner are of tetrahedral shape. 

\subsection{Figure's surface}

The boundary of the figure is defined by set of \textit{surface} equations for vertical and horizontal segment lines linking \textit{outer} nodes. After each mesh iteration new nodes are created and labeled as \textit{outer} or \textit{inner} ones according to surface equations. Moreover, the location of each node (i. e. on which exactly vertical, horizontal line or surface patch the node is lying) is also stored.  

\subsection{New elements creation}

New elements are created by a division of already existing elements. At the beginning of the routine, the surface of division mainly connects a new node born on the longest element edge with two other nodes belonging to that mesh element and one node from the divided edge. The procedure constitutes a 3D extension of the 2D mesh generation routine described already in \cite{kosinska_mesh2D}. However, during the routine a number of small elements is increasing, and the division of the longest bar is not anymore the optimal way of proceeding. That is why, before choosing an edge to the division volume of elements common to it is checked. The edge that will not produce new elements having its volume smaller than an assumed \textbf{critical volume} is chosen to be cut.  

\begin{figure}
\includegraphics[scale=0.55]{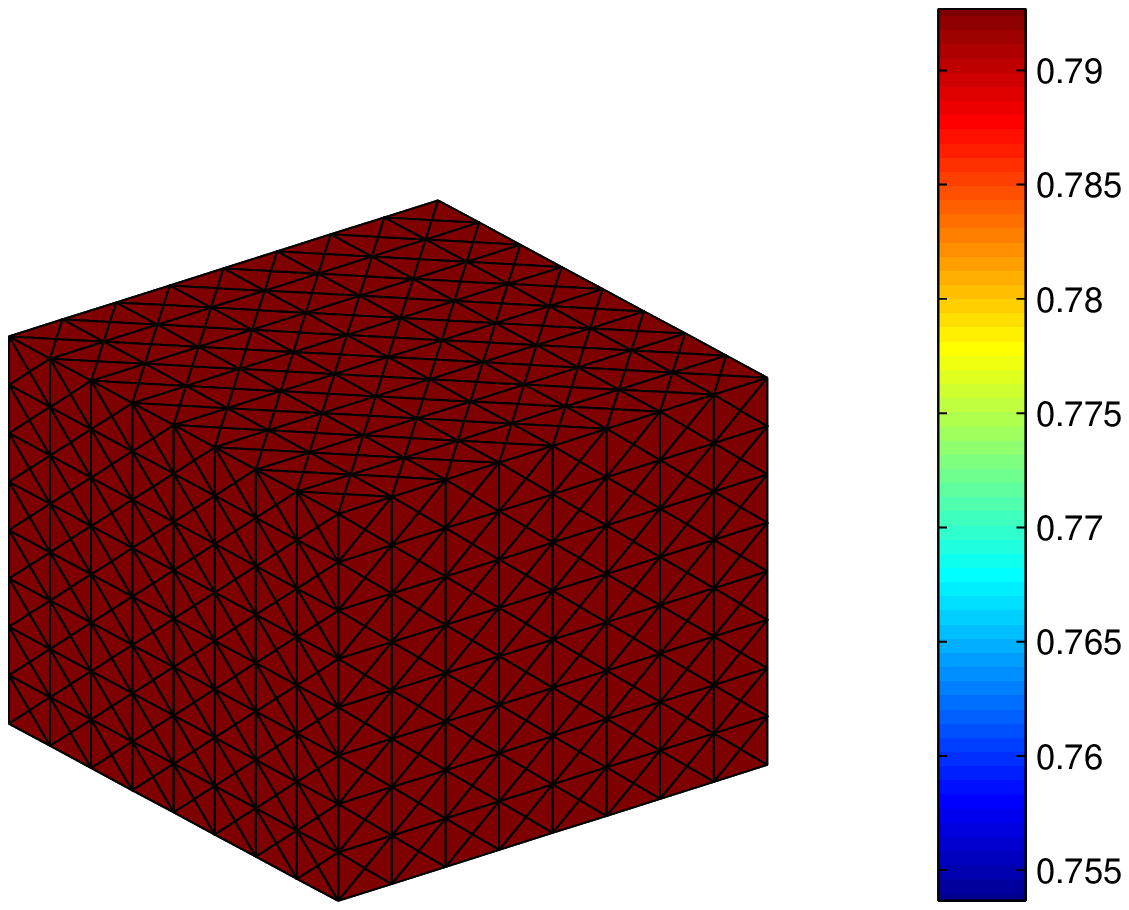}
\includegraphics[scale=0.55]{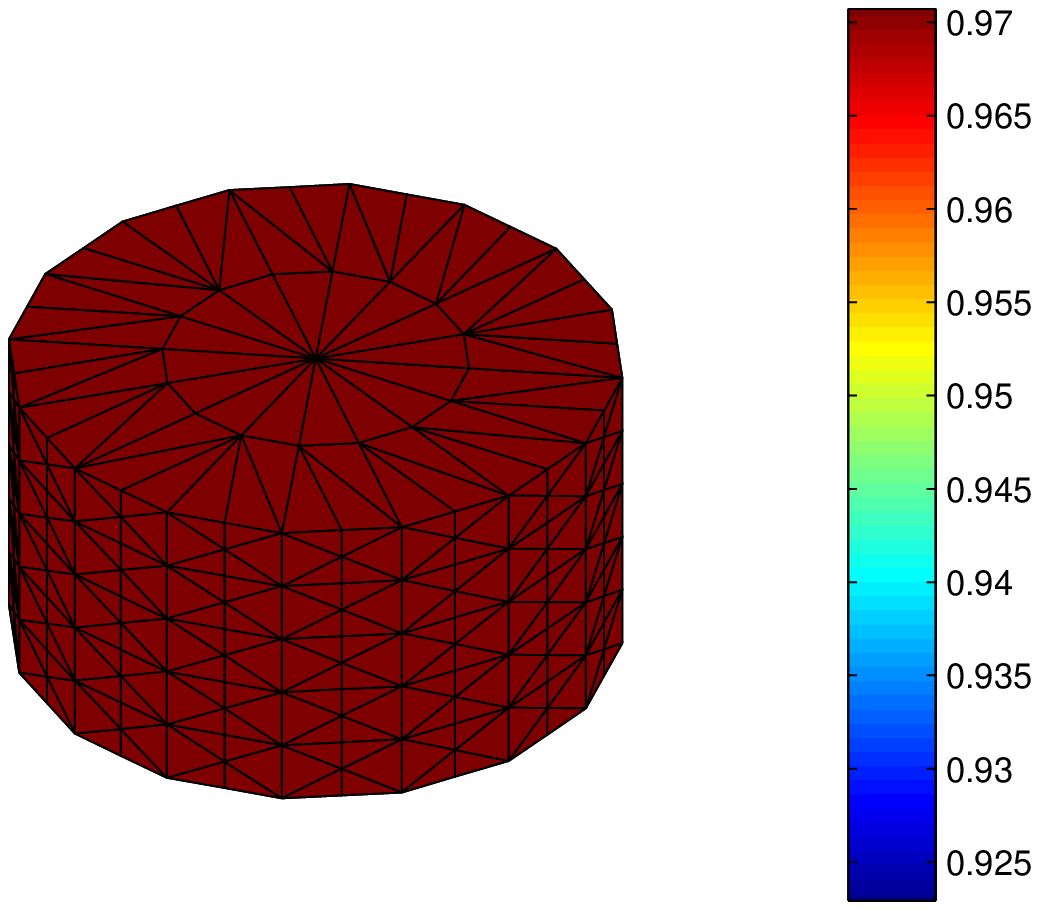}
\includegraphics[scale=0.55]{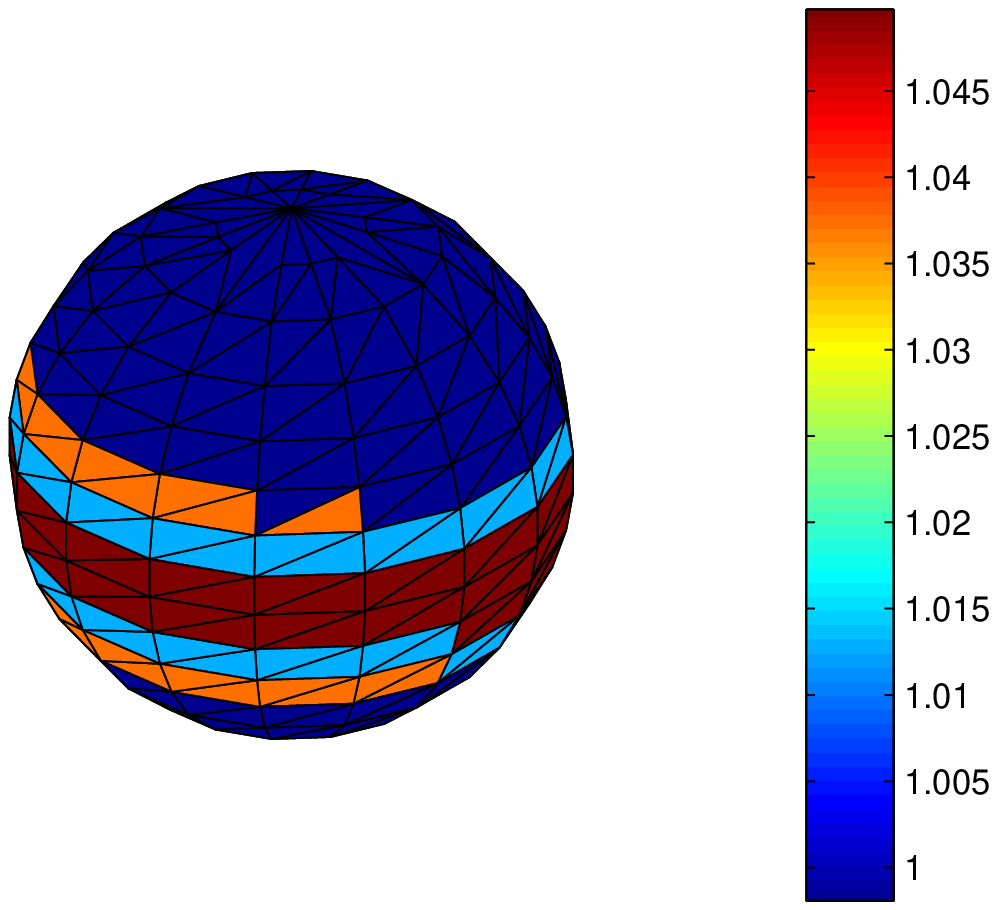}
\includegraphics[scale=0.55]{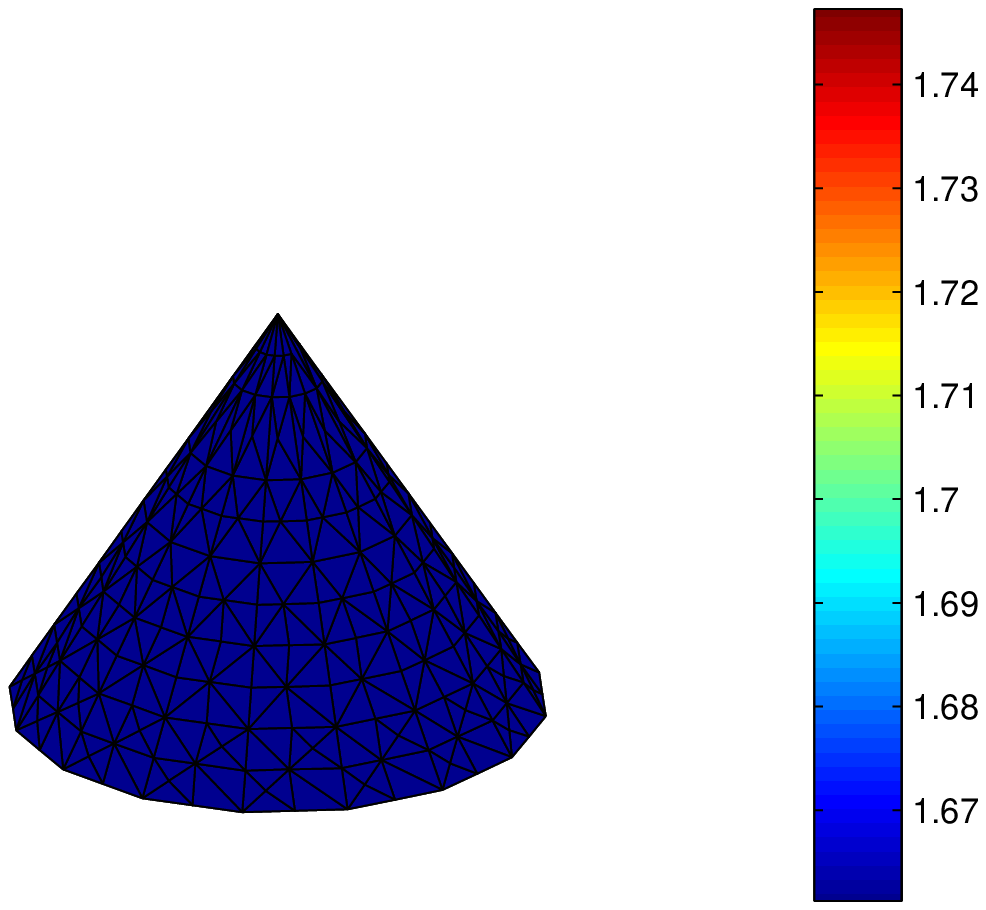}
\caption{The figure presents four regular shapes (cube, cylinder, sphere and cone) obtained with the non-optimized routine. Colorbars show variations in final elements volume.}
\label{regular}
\end{figure}

\subsection{Mesh optimization -- Metropolis algorithm}

The optimization is done with help of the Metropolis algorithm. The system energy is calculated as a sum of discrepancies between an element volume $V^e$ and assumed element volume $V_0 = h_0^3\sqrt{2}/12$ where $h_0$ denotes a prescribed length of the edge
\begin{equation}
E = \sum_e \left(V^e - V_0\right)^2.
\label{energy}
\end{equation}  
Thus the smaller a degeneracy from a designed volume distribution the more optimal state. The Metropolis routine starts from a nodal configuration given by described above procedure. The main point is to reach the optimal global configuration by ascertaining local optimal states. They arise from such a configuration of $i$-th node and its neighboring nodes which gives smaller energy $E^i$. This partial energy is calculated from the sum Eq.~(\ref{energy}) taken over elements containing the node of interest. To compute new positions for each node (giving new configuration) the following expression is put forward
\begin{eqnarray}
\displaystyle \mathbf{p}_{i,new} &=& \mathbf{p}_i 
- k_s\sum_{ij} \delta \mathbf{r}_{ij} \\
\delta \mathbf{r}_{ij} &=& \sum_j \left(\left|\mathbf{p}_i - \mathbf{p}_j\right| - h_0 \right)\frac{\mathbf{p}_i - \mathbf{p}_j}{\left|\mathbf{p}_i - \mathbf{p}_j\right|} 
\end{eqnarray}   
where $k_s$ denotes a shifting strength and $\mathbf{p}_i - \mathbf{p}_j$ is the length of $ij$ edge.
The value of $k_s$ determines the strength of a nodal shift and varies from 0 to 1. It is also worthy considering to choose its value as a random number from uniform distribution $U(0,1)$.
\begin{figure}
\includegraphics[scale=0.5]{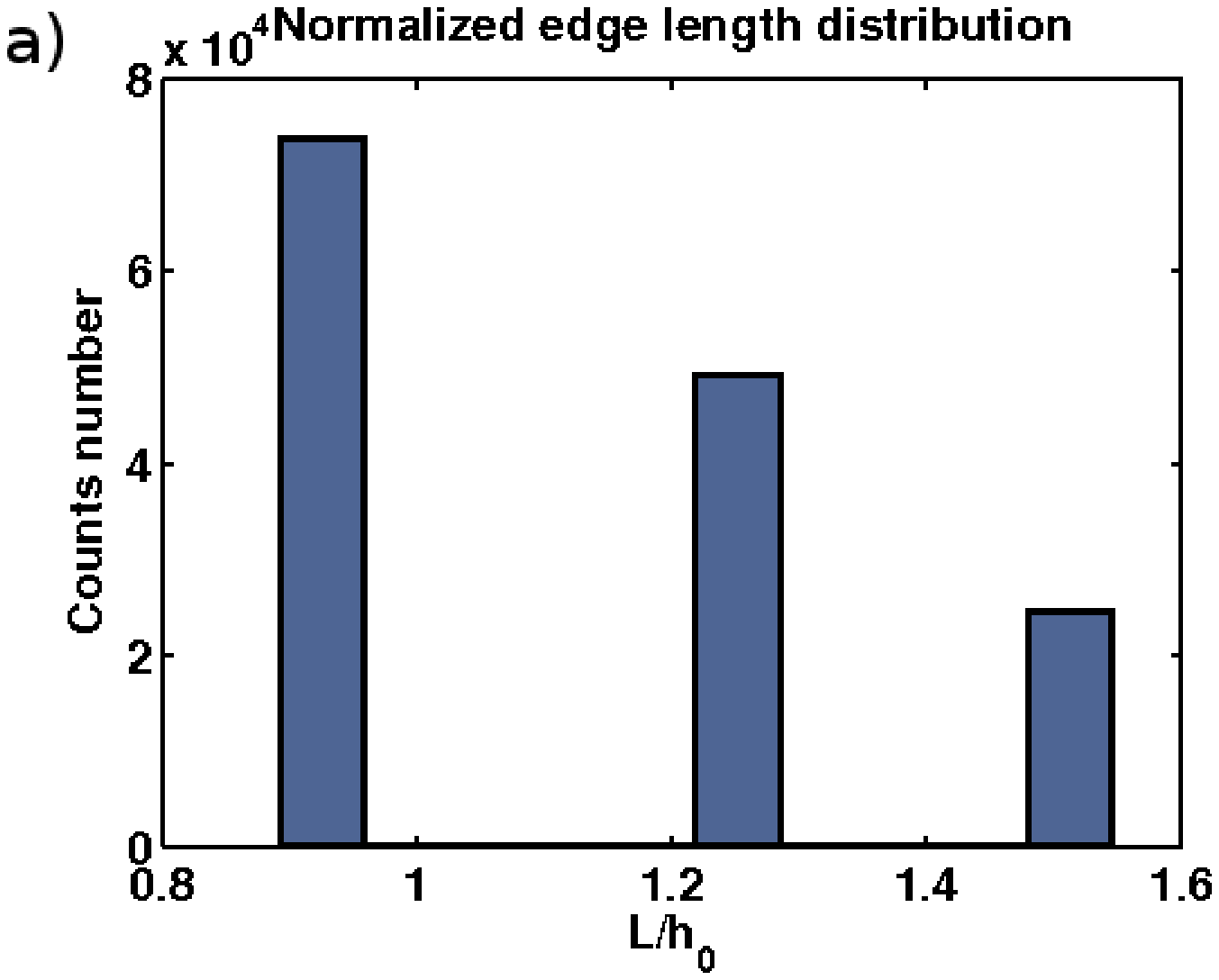}
\includegraphics[scale=0.5]{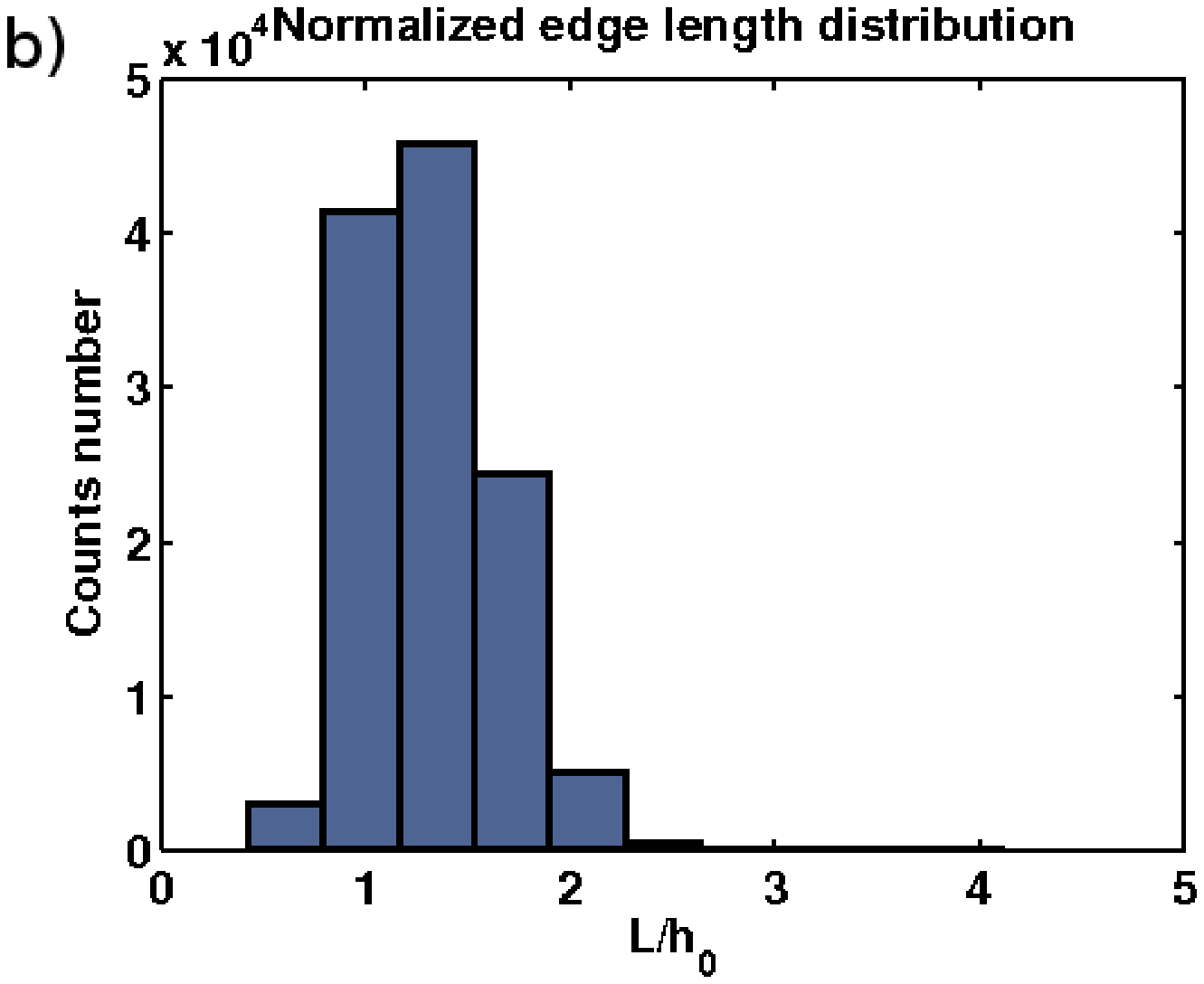}
\caption{The figure presents a distribution of a normalized mesh edge length $L/h_0$ for meshes in a cubic domain $[0, \pi]\times[0, \pi]\times[0, \pi]$ with a) unique elements created for $h_0 = 0.22$ and b) meshes optimized with the Metropolis algorithm with $h_0 = 0.2$.}
\label{length}
\end{figure}
Within the Metropolis routine the transition probability is calculated by the formula
\begin{equation}
P(p_i \to p_{i,new}) = e^{-\Delta E_i/k_BT}
\label{tran_prob}
\end{equation}
where $k_B$ is a Boltzmann constant (here set as 1), $T$ temperature and $\Delta E_i = E_{i,new} - E_{i}$ is a difference between energies of these two states. If a value of $P$ is greater than a random number from $U(0,1)$ new state is accepted. Otherwise the old one is preserved.\\ 
All above-described local Metropolis steps can lead to different global configurations. Therefore, for each division number this distribution of nodes which gives a lower energy of the total system should be kept. To find this, again the Metropolis rule is employed, but this time changes in the total energy of the whole system are examined. To estimate maximal temperature $T_{max}$ (in expression (\ref{tran_prob})) a range of changes in potential energy corresponding to current number of elements must be found. Moreover, in each global Metropolis step temperature might decrease according to $T=\eta T$ where parameter $\eta < 1$.   
\begin{figure}
\includegraphics[scale=0.5]{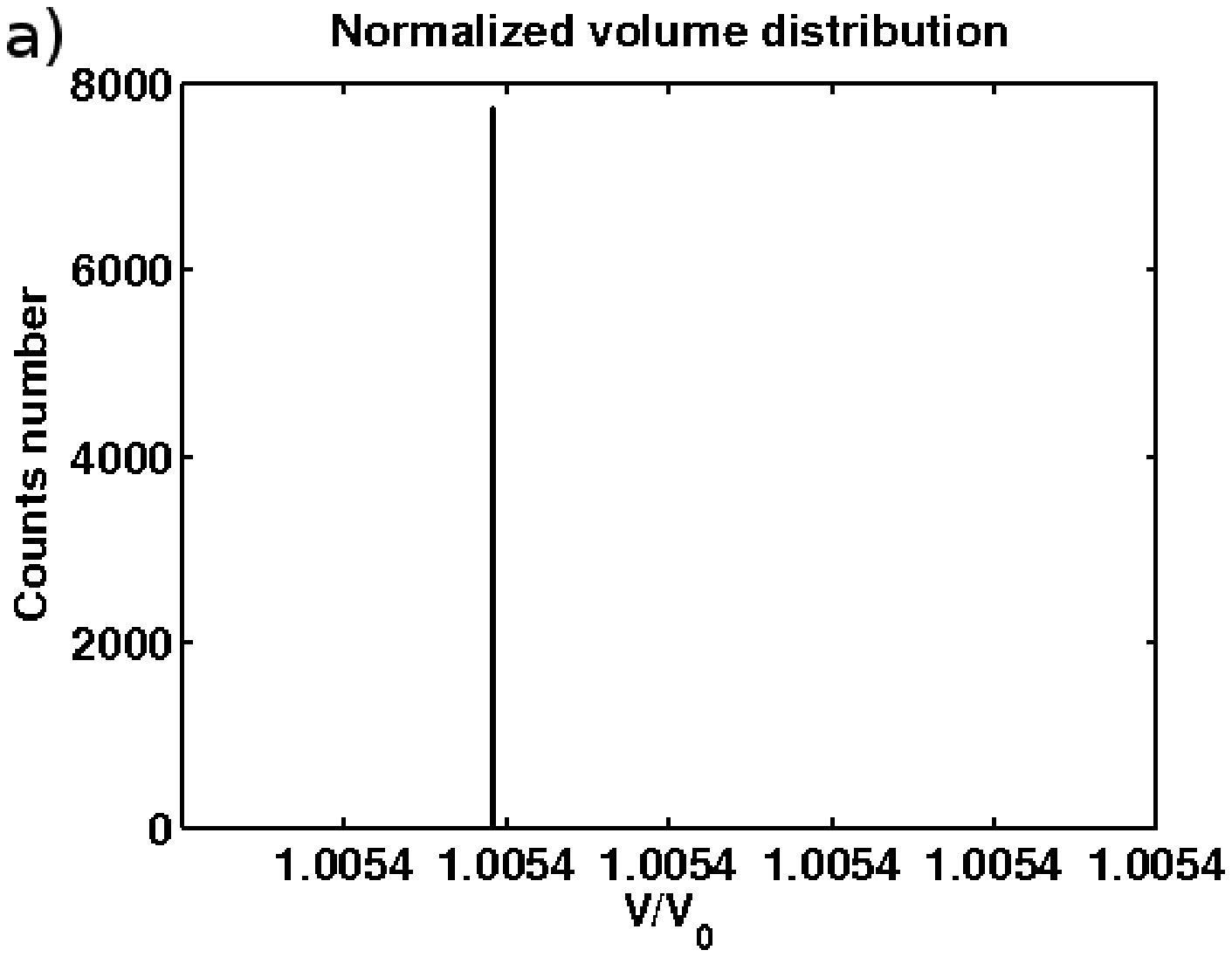}
\includegraphics[scale=0.5]{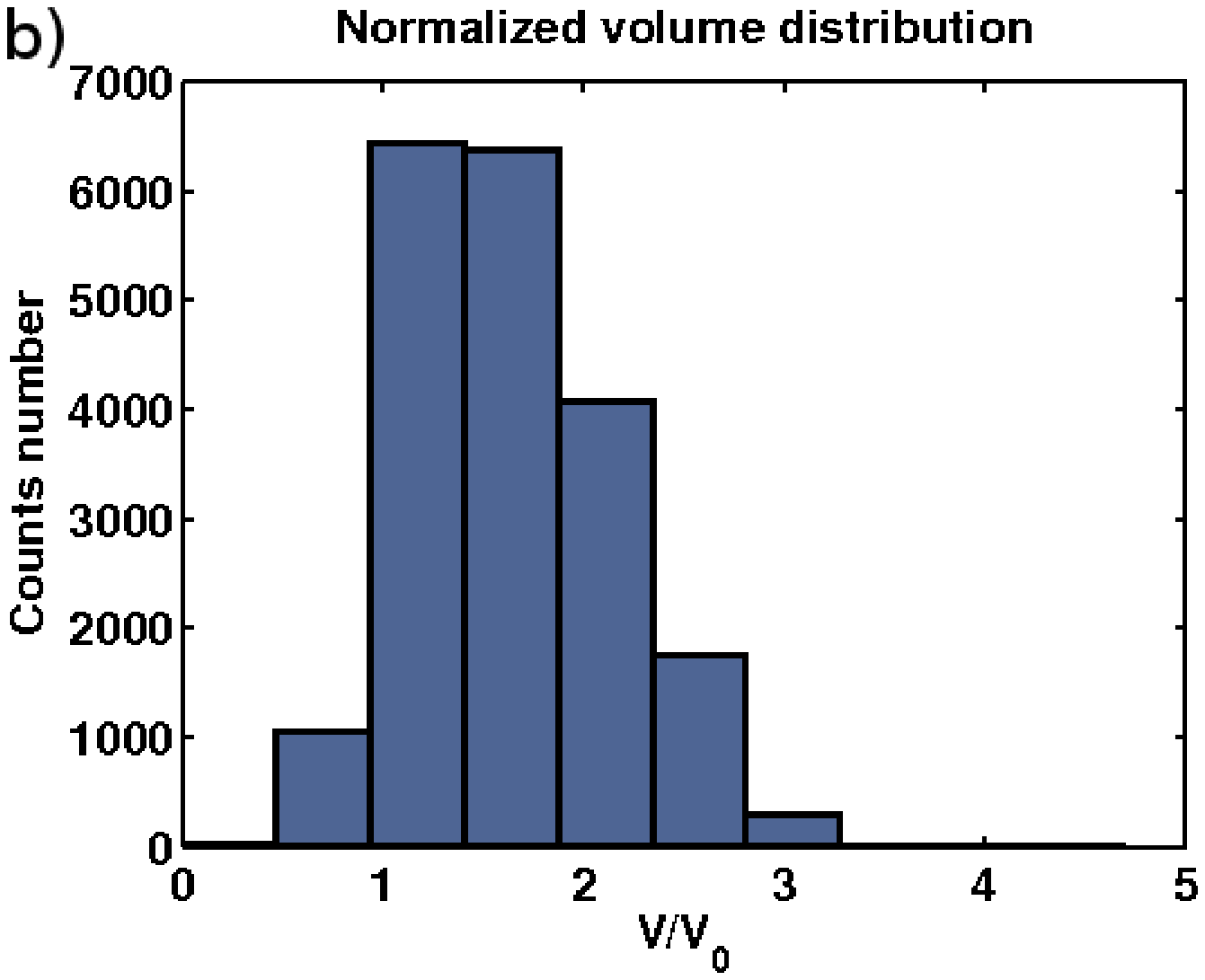}
\caption{The picture presents a distribution of a normalized mesh element volume $V/V_0$ for the same domain as in Fig.~\ref{length} and for a) unique elements; b) elements with $V_0 = 9\times10^{-4}$ optimized with the Metropolis algorithm.}
\label{volume}
\end{figure}

\subsection{Delaunay reconfiguration routine}

To improve mesh quality the following transformations are applied \cite{zienkiewicz}.
\begin{itemize}
\item
Three elements common to an edge are transformed to two elements, when one of the elements fails to satisfy the Delaunay criterion \cite{delaunay}.
\item
Four elements common to an edge are transformed to a new configuration of four elements, when one of the elements does not meet the Delaunay criterion \cite{delaunay}. The new pattern is chosen from two different possibilities \cite{zienkiewicz}. 
\end{itemize}  
Additionally, too small boundary elements could be destructed by a projection of its internal node to the center of the outer patch of element that is opposite to it. Such approach is justified in the case of boundary elements, however, in the case of internal ones leads to creation of so--called irregular nodes.  

\section{Results}

\subsection{The Laplace equation}

\begin{figure}
\includegraphics[scale=0.5]{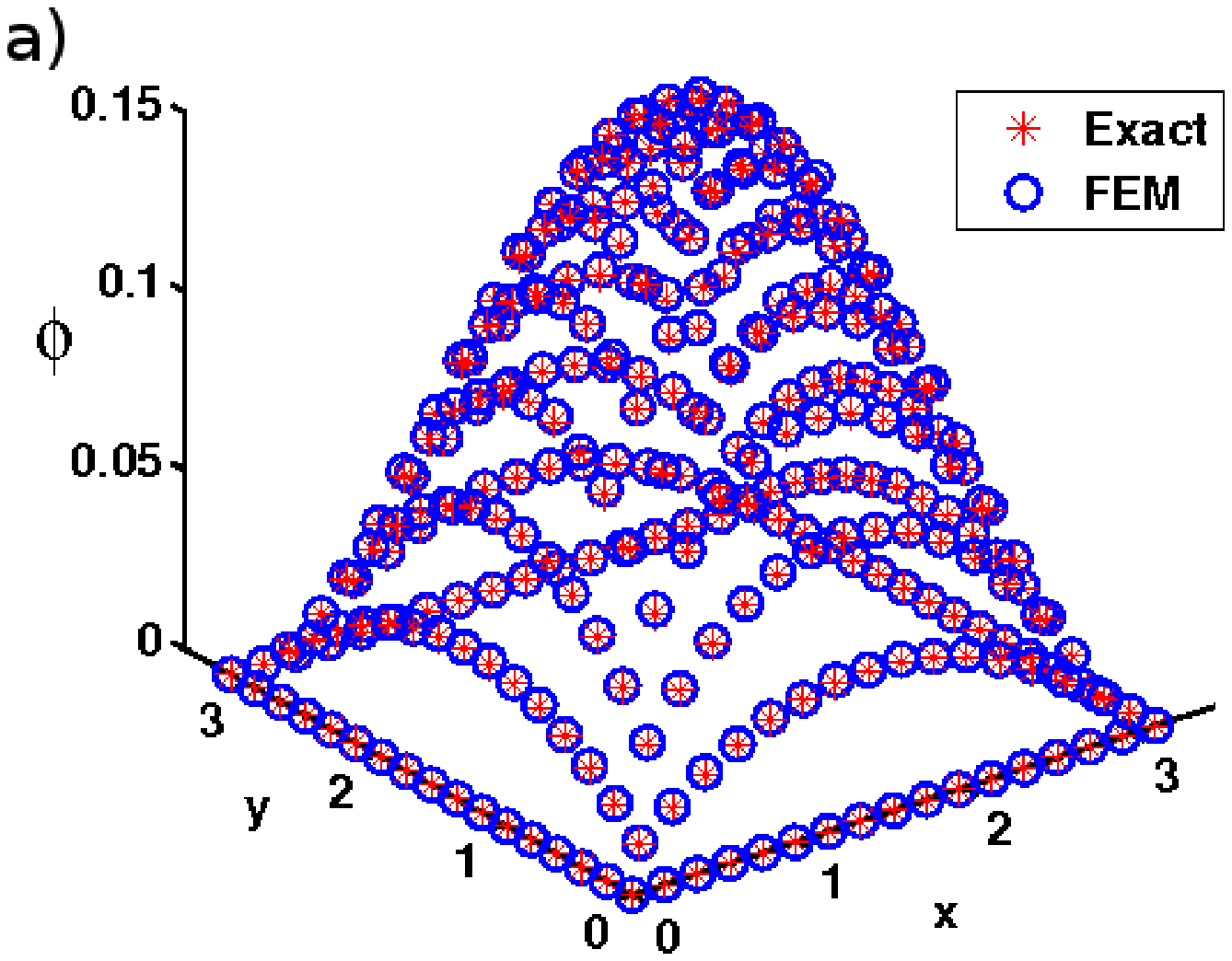}
\includegraphics[scale=0.5]{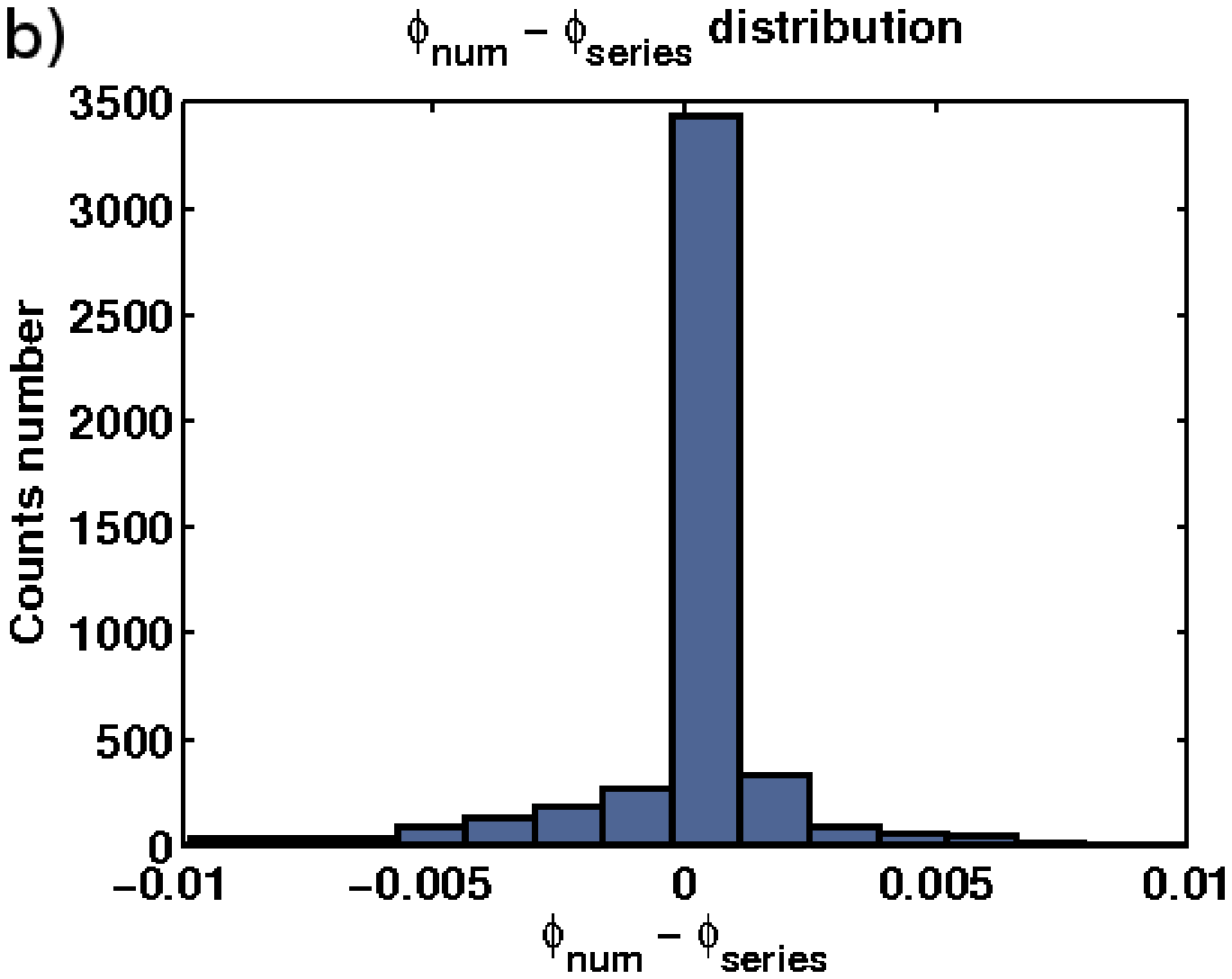}
\includegraphics[scale=0.5]{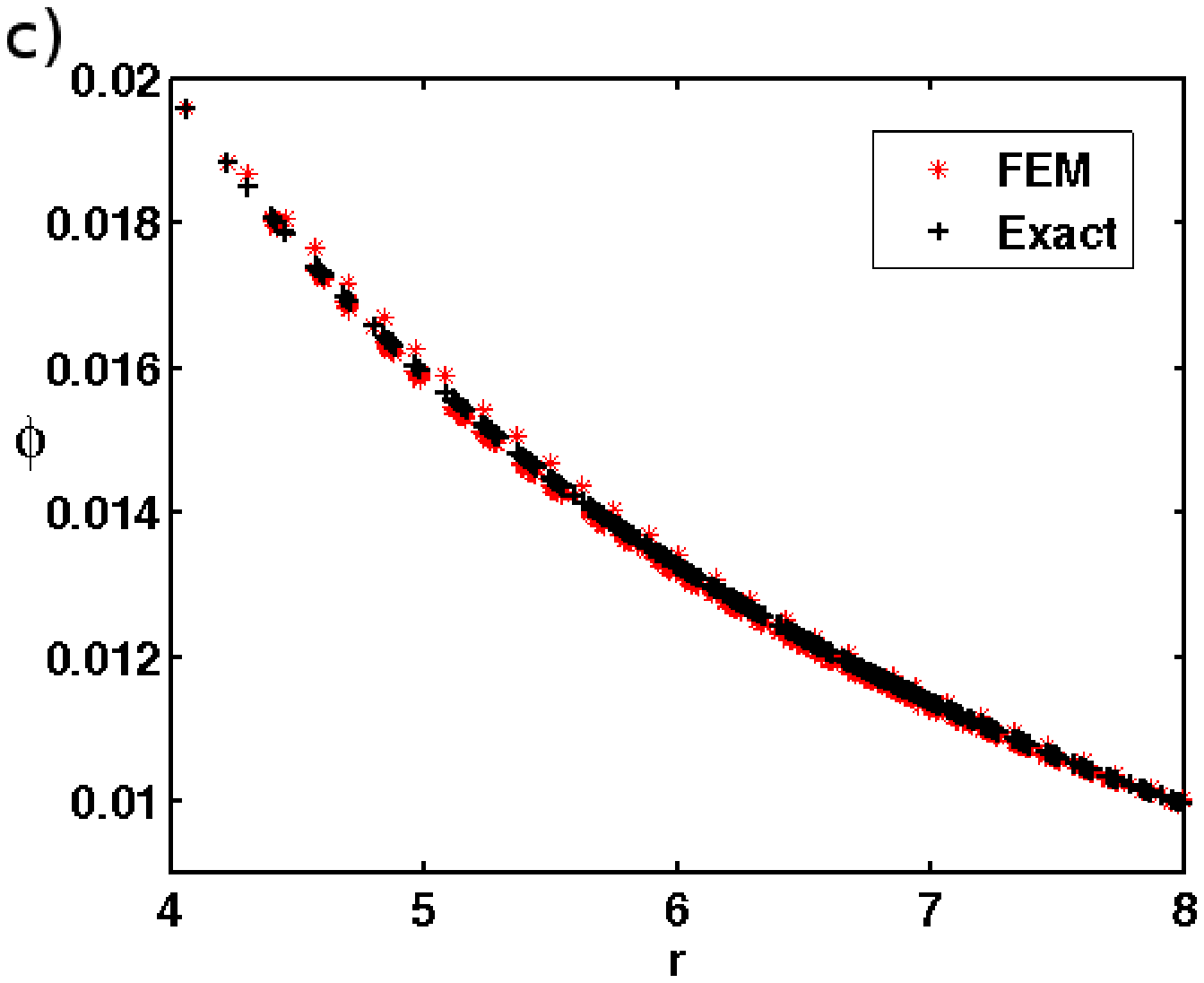}
\includegraphics[scale=0.5]{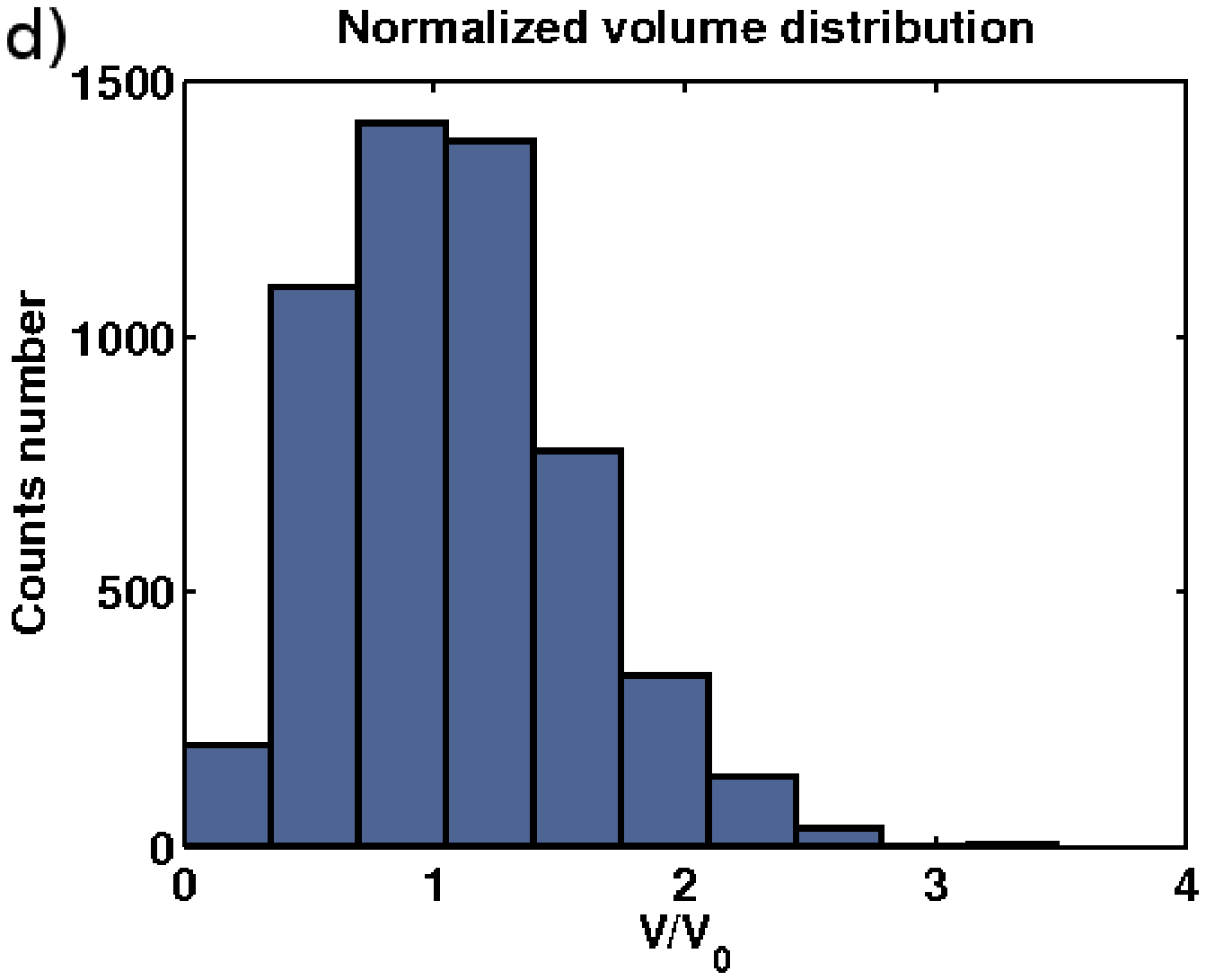}
\caption{The picture shows a) FEM approximation (linear) of the Laplace problem at the center of the cubic domain i. e. at $z=\pi/2$ (as described in Sec. 4) vs. the analytical result together with b) a distribution of differences between numerical and analytical solutions obtained for each node in $\Omega$ domain; c) FEM approximation (linear) vs. the exact solution of the Laplace equation for the spherical domain with the boundary values defined by putting an elementary charge outside the sphere in $[0, 0, 2\pi]$ b) the volume distribution in the spherical domain optimized with the Metropolis algorithm with elements of a prescribed volume $V_0$ equals $0.0075$.}
\label{laplace_cube}
\end{figure}

The accuracy of FEM approximation of the Laplace equation on different meshes were examined. Numerical results vs. analytical ones for cubic and spherical domain are presented in Fig.~(\ref{laplace_cube}). The relative difference between both analytical and numerical solutions has been calculated as 
\begin{equation}
(\phi_{num} - \phi_{anal})/\max(\phi_{anal}).
\label{difference_eq}
\end{equation} 
The Laplace equation has been solved for the cubic domain $[0, \pi]\times[0, \pi]\times[0, \pi]$ with potential function $\phi = 0$ everywhere on the boundary $\Gamma$ apart from one its side at $x = \pi$ where potential $\phi = 1$ and for the spherical domain with the boundary conditions imposed by putting an elementary charge outside the sphere in $[0, 0, 2\pi]$. The exact solutions for both considered cases are evaluated precisely in Sec. 4. FEM approximation has been computed for the ,,linear'' order of tetrahedron \cite{zienkiewicz}. However, the comparison between both orders of approximation i .e ,,linear'' and ,,quadratic'' for the uniform mesh with $V_{elem} = 0.0202$ has been performed. The formulas for higher orders of approximation i. e. quadratic and cubic can be found in \cite{zienkiewicz}. The results show that mean discrepancy between numerical and analytical solutions calculated according to Eq.~(\ref{difference_eq}) for the Laplace equation equal $-0.0061 \pm 0.0153$ in the linear case and $ 0.0004\pm 0.0082$ in quadratic approximation, respectively. Thus in further studies linear approximation will be used as being sufficiently accurate. In the case of cubic domain a mesh of unique element volume (non--optimized) has been applied in contrast to the spherical domain where mesh has been used after its enhancement with both the Metropolis algorithm and the Delaunay routine.  

\subsection{The diffusion equation}

\begin{figure}
\includegraphics[scale=0.5]{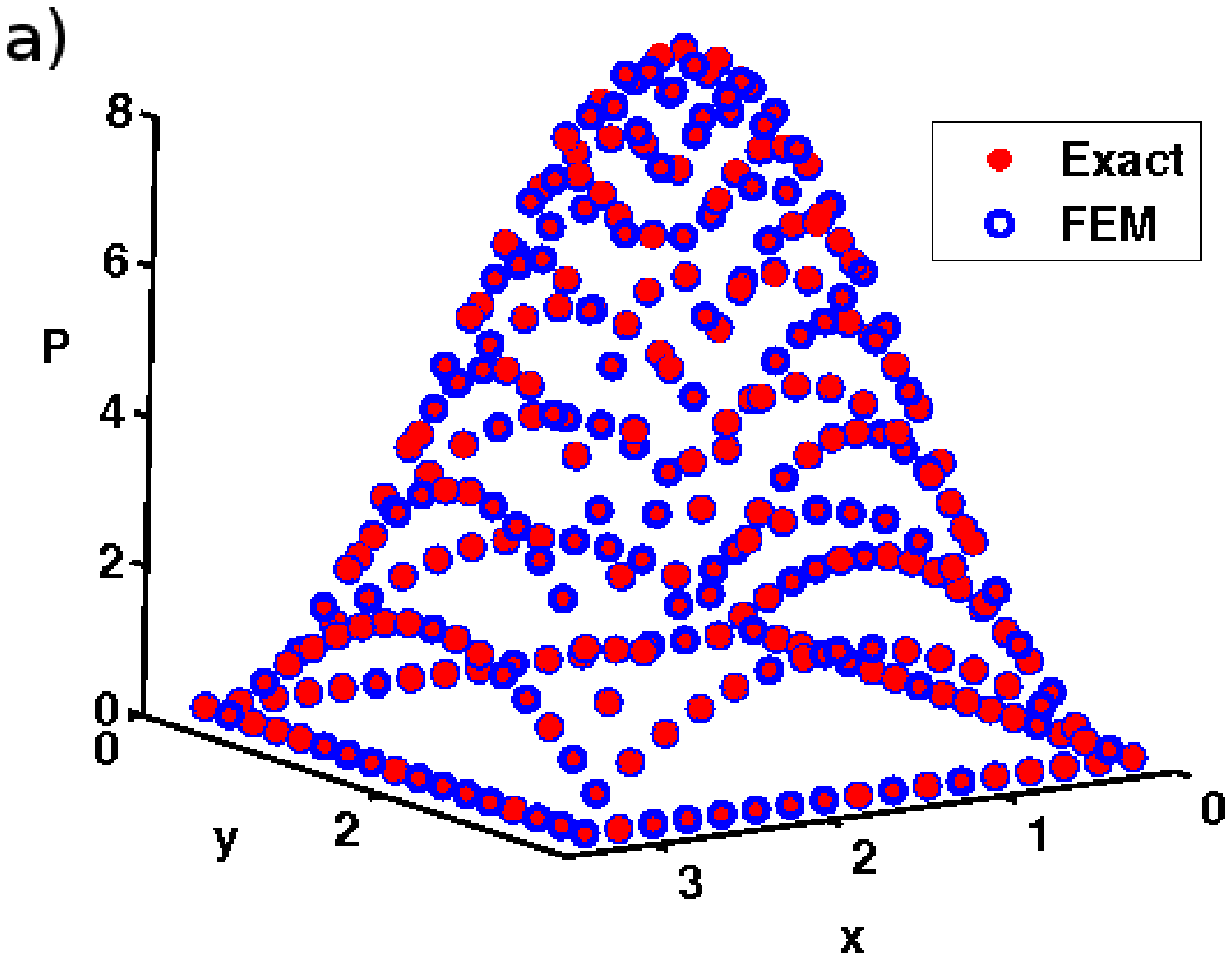}
\includegraphics[scale=0.5]{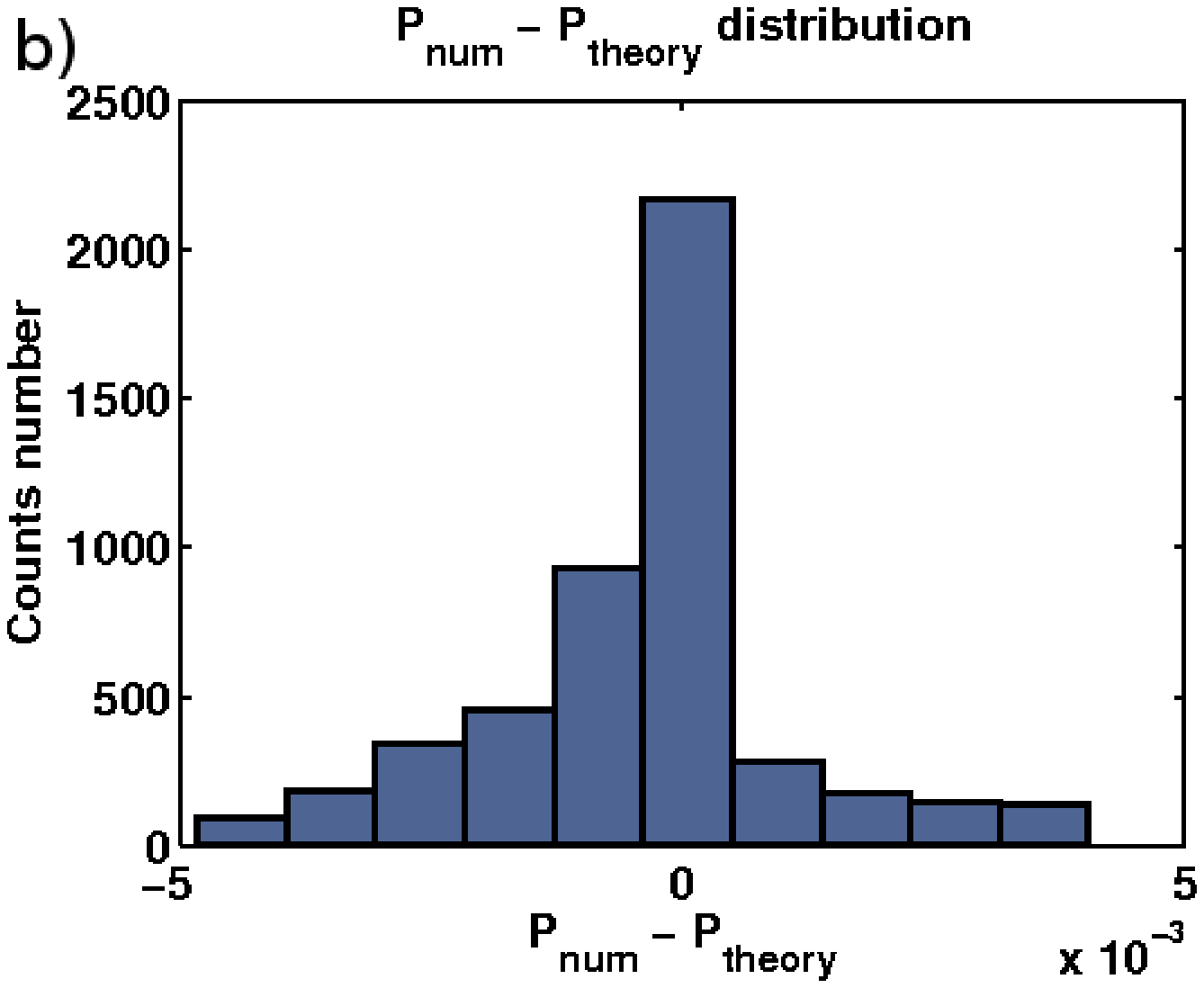}
\includegraphics[scale=0.5]{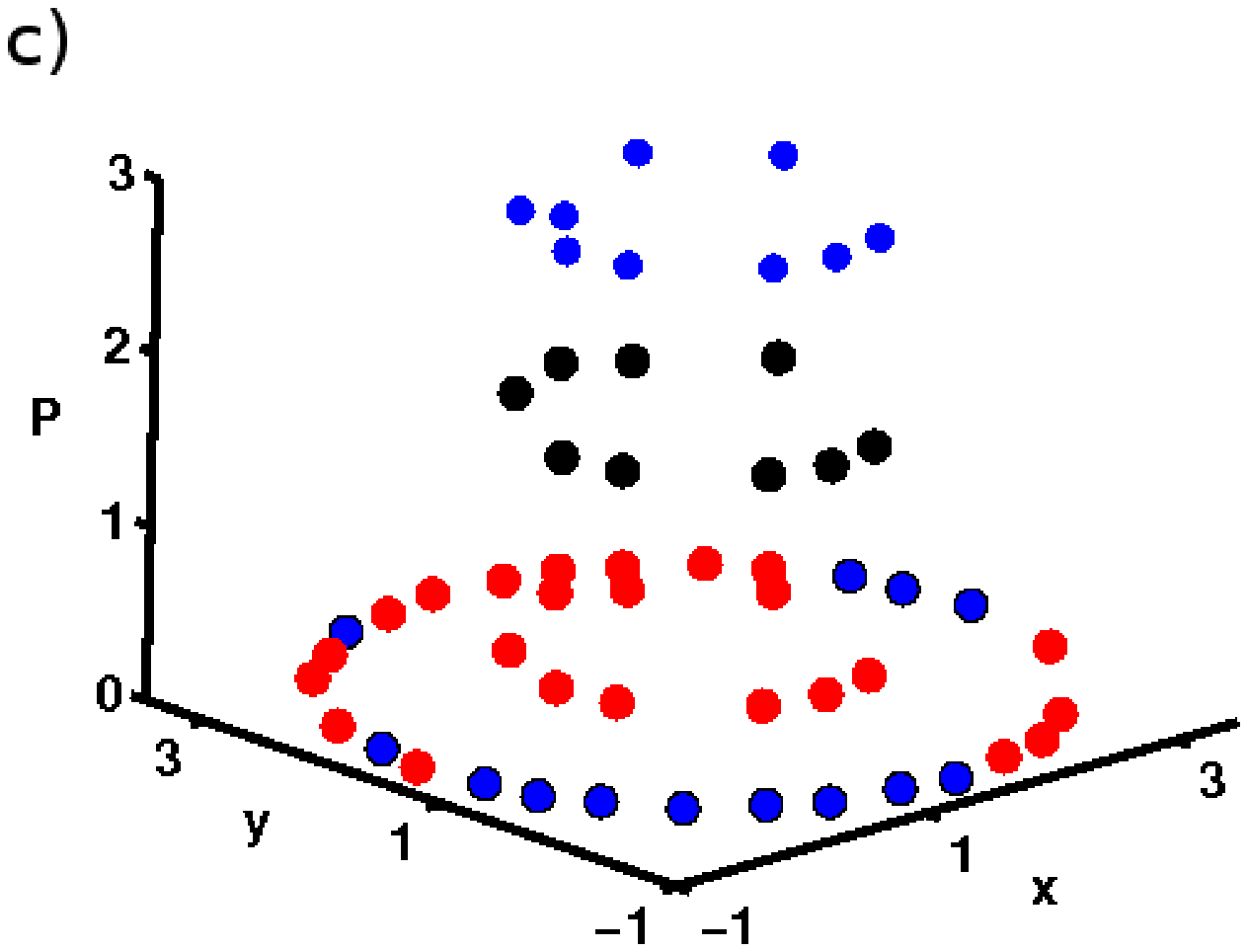}
\includegraphics[scale=0.5]{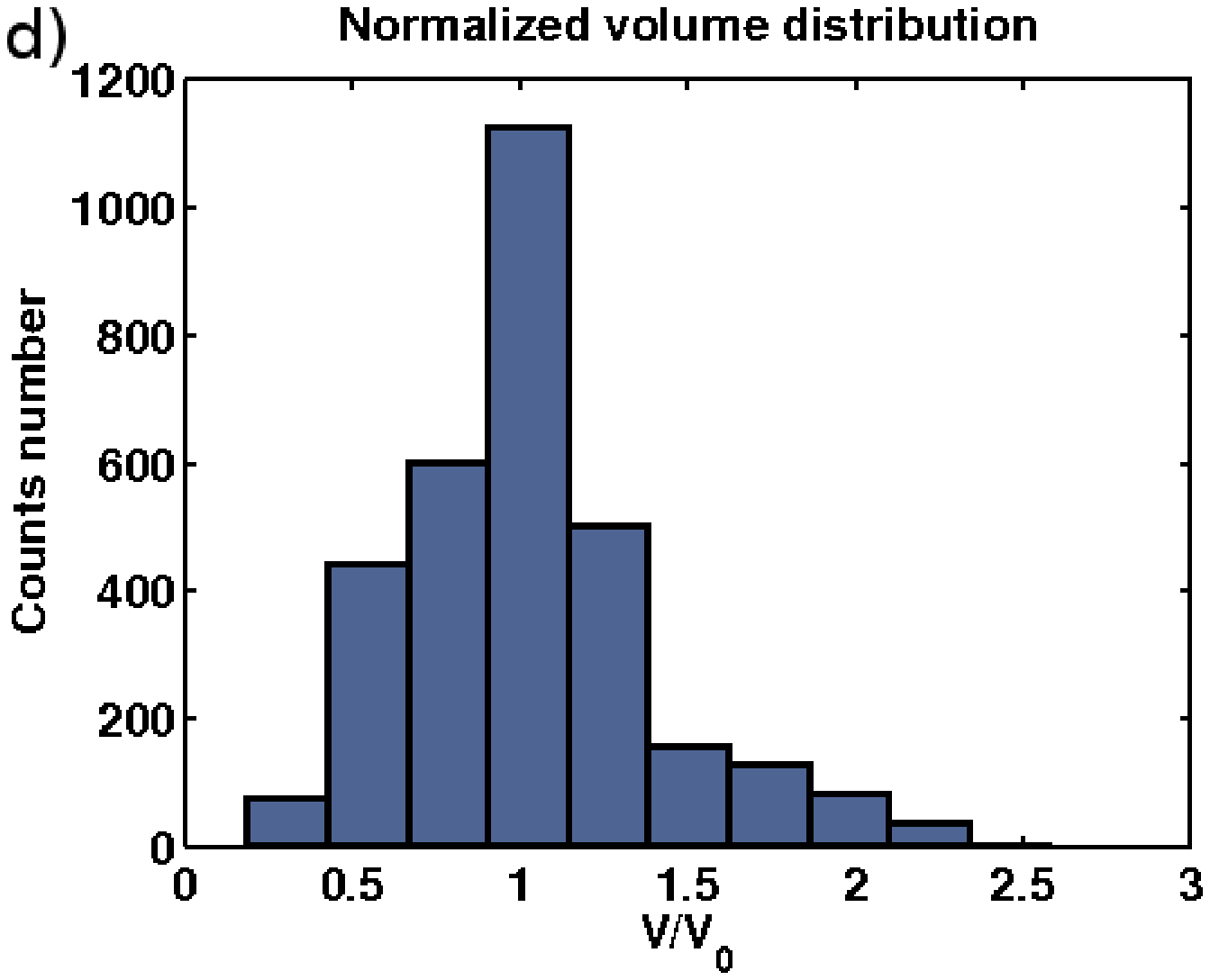}
\caption{The picture shows a) FEM approximation (linear) of the initial--boundary value problem in the center of the cubic domain i. e. at $z=\pi/2$ and at the time $t = 0.19$ with $\Delta t = 0.01$ [units] vs. the analytical result together with b) a distribution of differences (Eq.~\ref{difference_eq}) between numerical and analytical solutions obtained for each node in $\Omega$ domain; c) FEM approximation (linear) of the diffusion equation for the cylindrical domain at $z=\pi/2$ and at the following times: $t_0=0$ (blue), $t_{mid}=0.5$ (black) and $t_{end} = 1.0$ (red) with $\Delta t = 0.01$ [units]; d) volume profile of elements within the cylindrical domain where $V_0 = 0.015$ and $h_0 = 0.5$; mesh quality were enhanced with help of the Metropolis algorithm.}
\label{diffusion_pic}
\end{figure}

To test accuracy of discrete approximation in time the equation of diffusion (see Sec. 3) has been examined. To find $P(x,y,z,t)$ particles distribution at $t_{n}$ moment in time the Taylor expansion (see Eq.~\ref{discrete_in_time}) with $\beta=0$ has been applied. The diffusion equation with the following initial condition $g(\cdot, 0) = -x(x-\pi)y(y-\pi)z(z-\pi)$ for the cubic domain and $g(\cdot, 0) = |(r - R_0)z(z - \pi)|$ where $R_0$ denotes a radius of the cylindrical domain has been solved. The boundary value of the $P(x,y,z,t)$ is set as 0. Results for both domains: cubic (with uniform elements -- see Fig.~\ref{volume}a) and cylindrical (see Fig.~\ref{regular}), this one tuned to the designed element volume with the Metropolis recipe, are shown in Fig.~\ref{diffusion_pic}. In the case of cylindrical domain mean values of the ratio $P(\vec{x},t_{i})/P(\vec{x}, t_{i+10})$ calculated at each point of domain for times $i = 0, 10, \dots, 100$ have the average value $1.2699 \pm 0.0049$.  

\subsection{The electrodiffusion equation}

\begin{figure}
\includegraphics[scale=0.5]{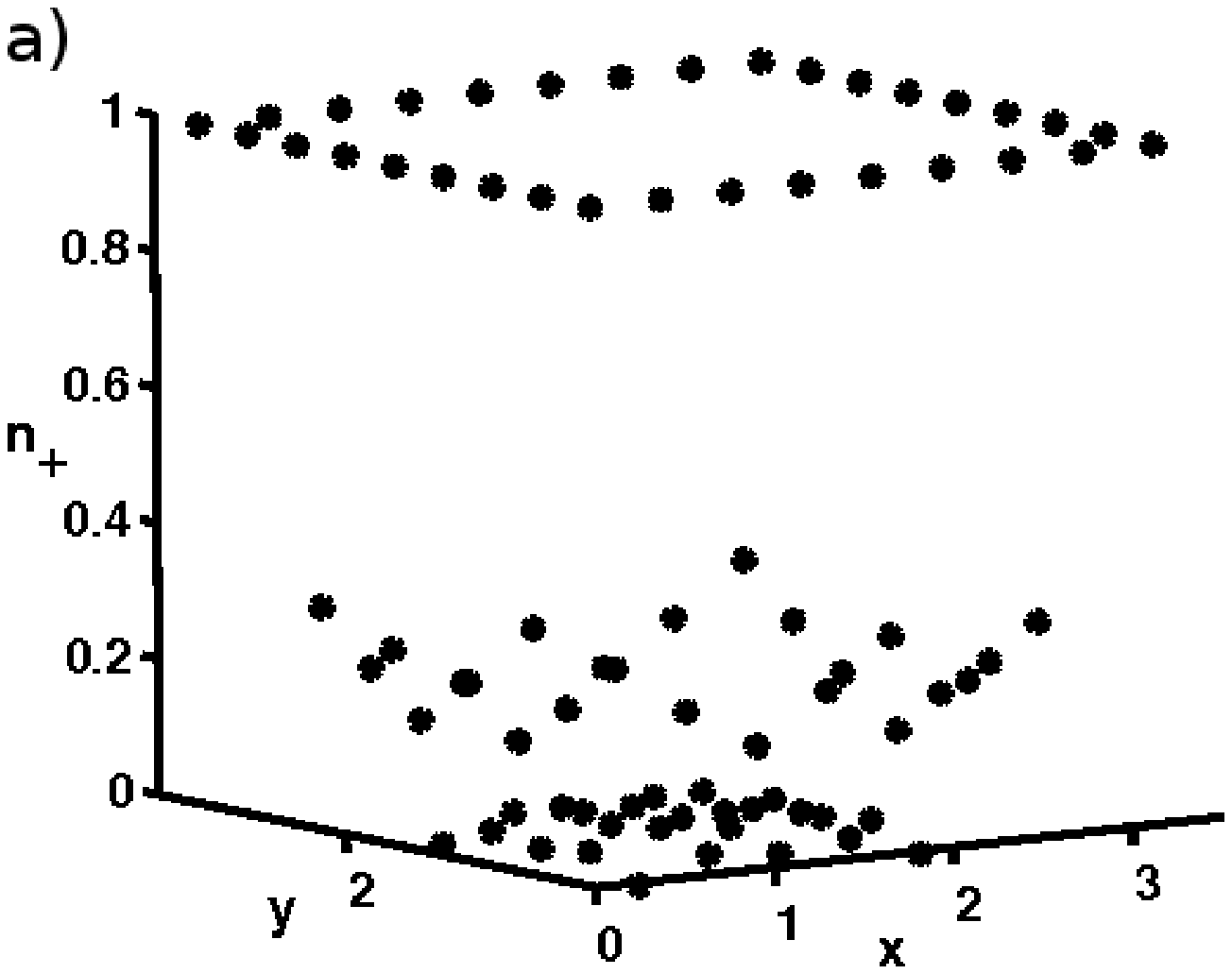}
\includegraphics[scale=0.5]{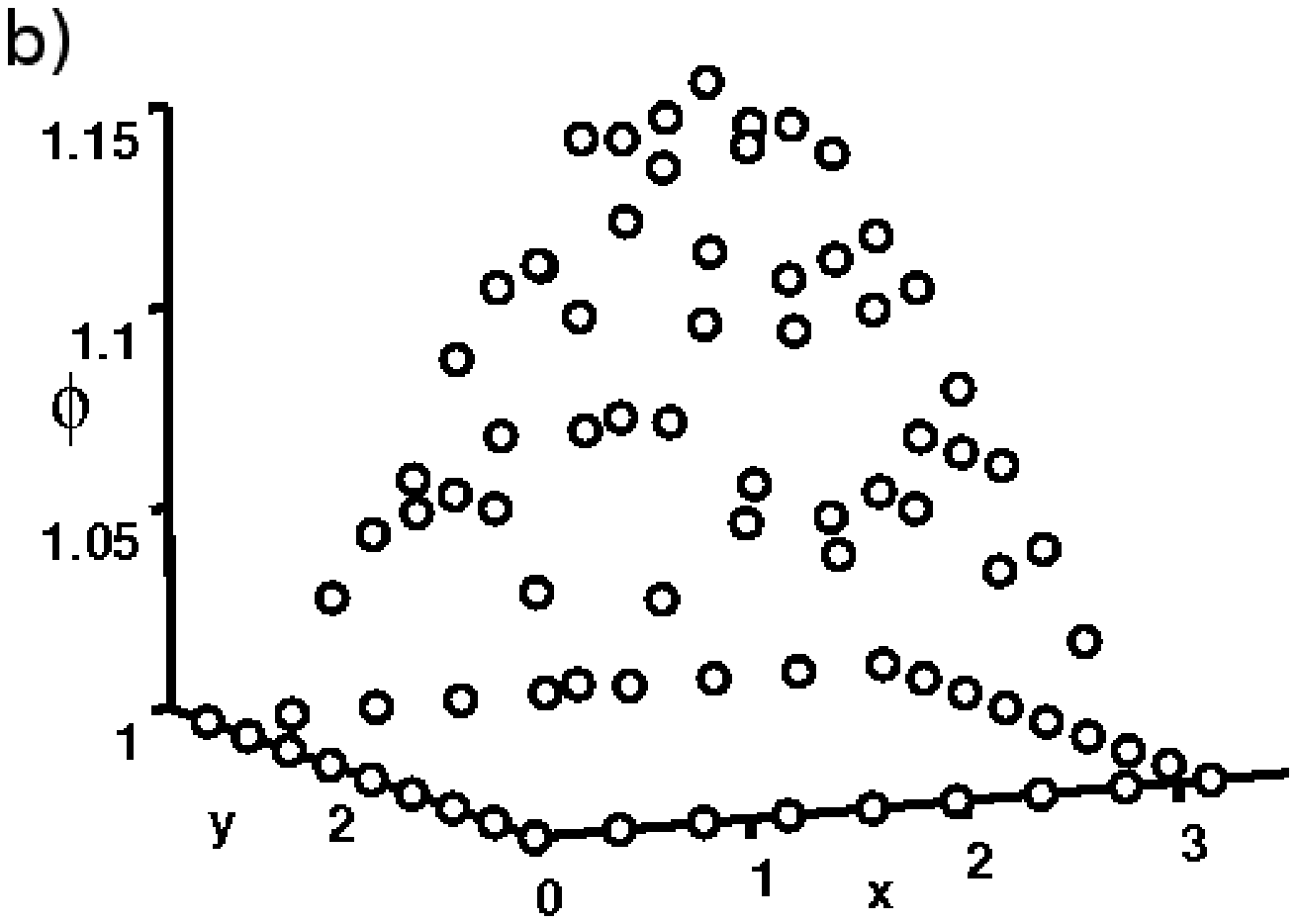}
\caption{The picture shows FEM approximation (linear) of the system of electrodiffusion equations being the boundary value problem. The solutions are depicted in the center of the cubic domain i. e. at $z=\pi/2$ and at the time $T = 0.39$ with the values of parameters $\Delta t = 0.01$ [units], $k_{+} = k_{-} = 0.05$, $D_{+} = D_{-} = 0.05$ [units] a) the distribution of cations $n_{+}$; b) the profile of the potential $\phi$; computations have been performed on the uniform mesh with the volume of tetrahedron $V_0 = 0.01$ and the element size $h_0 = 0.44$.}
\label{electrodiffusion_pic}
\end{figure}

The system of coupled equations describing process of electrodiffusion (\ref{electrodiffusion}) written in terms of the FEM method (see Eq.~(\ref{pnp_final})) with the following values of constants: $k_{+} = k_{-} = 0.05, D_{+} = D_{-} = 0.05$ and the time step equals $0.01$ has been numerically solved by using the Newton's algorithm. The boundary values of $n_{+}, n_{-}, \phi$ are set as 1 at $x = 0$, $x = \pi$, $y = 0$, $y = \pi$, and $z = 0$ and equal 2 at $z = \pi$. An initial guess of $n_+, n_-$ and $\phi$ distributions has been chosen as 0 everywhere in the domain apart from its boundaries. The system of equations has been computed up to the final time $T = 0.39$. Fig.~\ref{electrodiffusion_pic} presents obtained profiles of cations $n_{+}$ and the potential $\phi$ at the center of the domain i. e. at $z = \pi/2$. There is no visible difference between cations $n_{+}$ and anions $n_{-}$ distributions so the latter is not shown.
The maximum of $n_{+, i+1} - n_{+, i}$ is decaying from 0.023 for $i = 0$ ($t = 0$) to 0.0093 for $i = 37$ ($t = 0.37$). The maximum of difference between $n_+$ and $n_-$ distributions computed at each node equals $1.3e-09$. Employing physical notion it means that the system of charged particles is electroneutral. Moreover, the system of particles is tending to its stationary state. 

\begin{figure}
\includegraphics[scale=0.5]{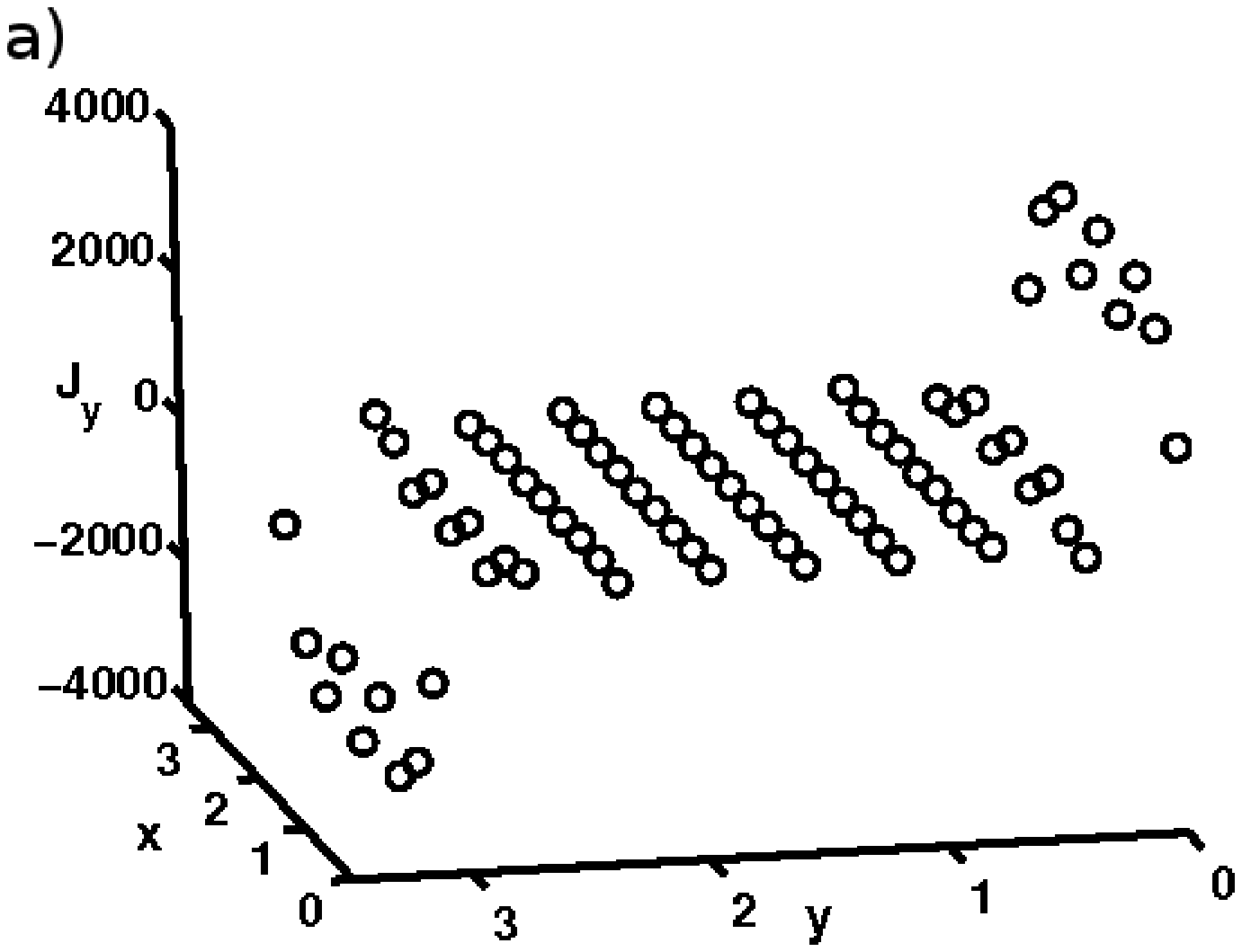}
\includegraphics[scale=0.5]{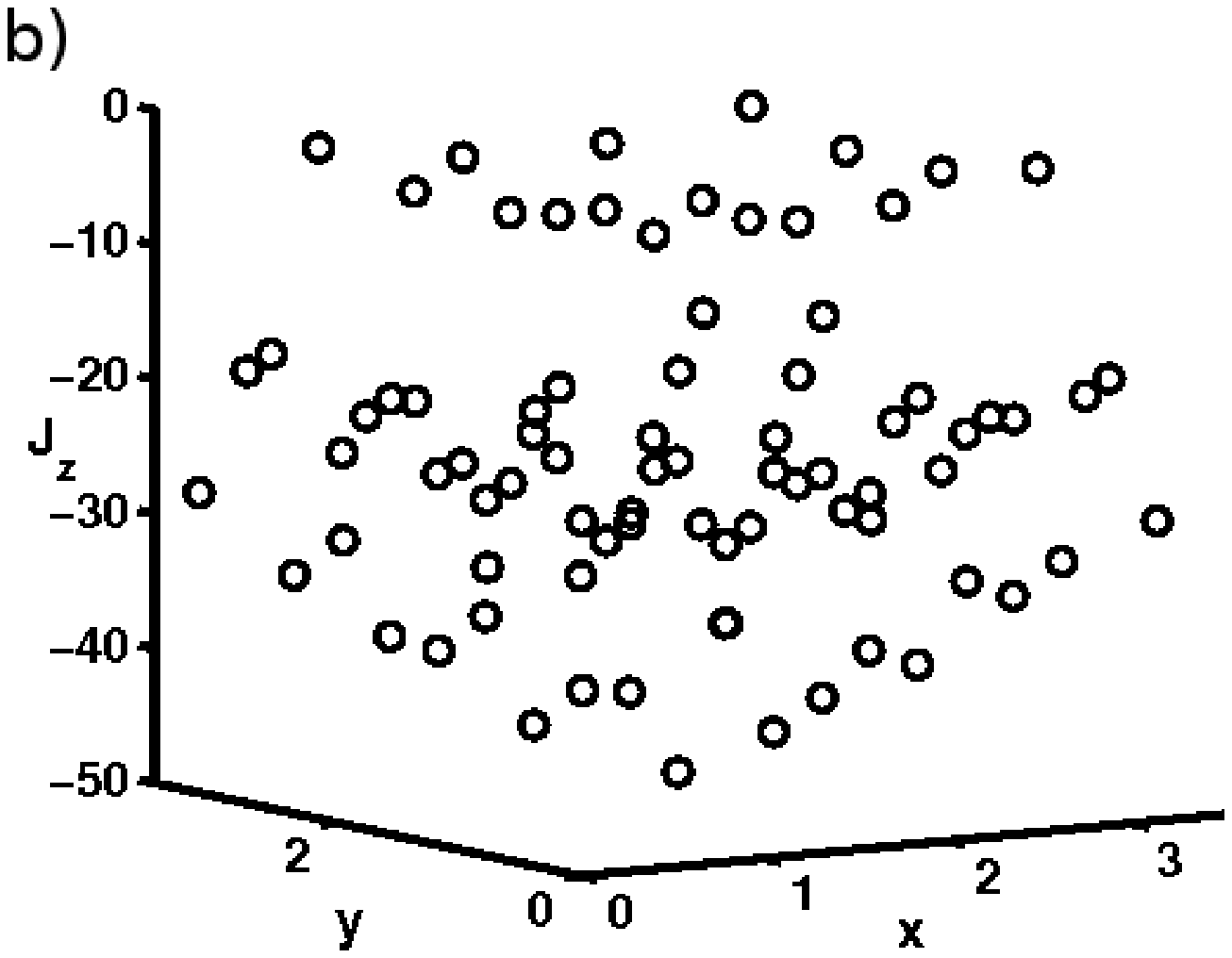}
\caption{The picture presents FEM approximation (linear) to the flux components obtained for the electrodiffusional problem. The solutions are shown at the center of the cubic domain i. e. at $z=\pi/2$ and at the time $T = 0.39$. Computations have been done with the following values of parameters: the time step $\Delta t = 0.01$ [units], nonlinear term multipliers $k_{+} = k_{-} = 0.05$ [units], diffusion coefficients $D_{+} = D_{-} = 0.05$ [units]. This figure shows  approximated solutions for flux components for cations $n_+$ in the $y$ direction a) and in the $z$ direction b). The computations have been performed on the uniform mesh with the volume of tetrahedron $V_0 = 0.01$ and the element size $h_0 = 0.44$.}
\label{electrodiffusion_current}
\end{figure}

Additionally, components $j_x, j_y$ and $j_z$ of the total flux of $n_+$ particles flowing through the domain $\Omega$ have been computed. They are shown in Fig.~\ref{electrodiffusion_current}. Presence of a difference in an amount of $n_{+}$ particles at the both sides of $\Omega$ in the $z$--direction i. e. at $z=0$ and $z=\pi$ causes a non--zero flow along $z$ axis whereas a lack of such a difference in the two other directions i. e. $x$ and $y$ leads to the vanishing flows$j_x$ and $j_y$ in the center of the domain.  

\section{Conclusions}

The presented software offers a 3D mesh generation routine as well as its further application to the 3D electrodiffusional problem.\\ 
The proposed mesh generator offers a confident way to creature a quite uniform mesh built with elements having desired volume. Mesh elements have been adjusted to assumed sizes by making use of both the Metropolis algorithm and the Delaunay criterion. Mesh quality depicted in histograms occurs to be fairly satisfactory. Moreover, goodness of obtained meshes together with robustness of their applications to the Finite Element Method have been also tested by solving the 3D Laplace problem and the 3D diffusion equation on them. Comparison between these numerical solutions and analytical results shows very good agreement.\\     
To find solutions to a nonlinear problem defined by a system of coupled equations describing electrodiffusion the FEM approach and the Newton method have been jointly applied. Analysis of obtained results confirms usefulness of the presented solver to deal with nonlinear differential problems.

\end{document}